\documentclass[oldversion,fleqn]{aa}
\usepackage{natbib}
\bibpunct{(}{)}{;}{a}{}{,} % to follow A&A style
\usepackage{amsmath,amssymb}

\usepackage{graphicx}

\usepackage{txfonts}
\newcommand{\moped}{\small MOPED\normalsize}
\newcommand{\Max}{MA$\chi$}
\newcommand{\caii}{Ca\small II \normalsize}
\newcommand{\mgi}{ Mg\small{I}\normalsize}

\begin{document}
   \title{Stellar atmosphere parameters with MA$\chi$, a MAssive compression of $\chi^2$ for spectral fitting}

   %\subtitle{I. Overviewing the $\kappa$-mechanism}

   \author{P. Jofr\'e
          \inst{1}
          \and
          B. Panter\inst{2}  \and C.~J. Hansen\inst{3} \and A. Weiss\inst{1}}

%\fnmsep\thanks{Just to show the usage of the elements in the author field}

   \offprints{P. Jofr\'e}
\institute{
   Max-Planck-Institut f\"ur Astrophysik, Karl-Schwarzschild-Str. 1, 85741 Garching, Germany\\
  \email{jofre@mpa-garching.mpg.de}
  \and
  Institute for Astronomy, University of Edinburgh, Royal Observatory, Blackford Hill, Edinburgh EH9 3HJ
  \and
  European South Observatory (ESO),  Karl-Schwarzschild-Str. 2, 85748 Garching, Germany
  }

\authorrunning{Jofr\'e et al.}
\titlerunning{Stellar atmosphere parameters with MA$\chi$}
   \date{}%Recived, acceptet

%\abstract{}{}{}{}{}
% 5 {} token are mandatory

  \abstract
{MA$\chi$ is a new tool to estimate parameters from stellar spectra. It is based on the maximum likelihood method, with the likelihood compressed in a way that the information stored in the spectral fluxes is conserved. The compressed data are given by the size of the number of parameters, rather than by the number of flux points. The optimum speed-up reached by the compression is the ratio of the data set to the number of parameters. The method has been tested on a sample of low-resolution spectra from the Sloan Extension for Galactic Understanding and Exploration (SEGUE) survey for the estimate of metallicity, effective temperature and surface gravity, with accuracies of  0.24 dex, 130K and 0.5 dex, respectively. Our stellar parameters and those recovered by the SEGUE Stellar Parameter Pipeline agree reasonably well. A small sample of high-resolution VLT-UVES spectra were also used to test the method and the results have been compared to a more classical approach. The speed and multi-resolution capability of MA$\chi$ combined with its performance compared with other methods indicates that it will be a useful tool for the analysis of upcoming spectral surveys.
}{}


\maketitle
%
%________________________________________________________________

\section{Introduction}
\label{introduction}
The astronomical community has conducted many massive surveys of the Universe, and many more are either ongoing or planned. Recently completed is the Sloan Digital Sky Survey \citep[SDSS,][]{sdss}, notable for assembling in a consistent manner the positions, photometry and spectra of millions of galaxies, quasars and stars, and covering a significant fraction of the sky. The size of such a survey allows us to answer many questions about the structure and evolution of our universe. More locally, massive surveys of stars have been undertaken to reveal their properties, for example the Geneva-Copenhagen Survey for the solar neighborhood \citep{geneva_copenhagen} and the ELODIE library \citep{elodie}. These surveys provide a copious amount of stellar photometry and spectra from the solar neighborhood, broadening the range of stellar types studied. The SEGUE catalog -- Sloan Extension for Galactic Understanding and Exploration  \citep{Yanny09} -- a component of SDSS, contains additional imaging data at lower galactic latitudes, to better explore the interface between the disk and halo population.\\
By statistically analyzing the properties of these stars such as chemical abundances, velocities, distances, etc, it has been possible to match the structure and evolution of the Milky Way to the current generation of galaxy formation models \citep{juric08, ivecic08, bond09, wyse}. \\

Studies of large samples of stars of our galaxy are crucial tests for the theory of the structure and formation of spiral galaxies. Even though the statistics given by SEGUE or the Geneva-Copenhagen survey agree with the models and simulations of galaxy formation, there is some contention over whether the accuracies of measurements adequately support the conclusions. New surveys such as RAVE \citep{rave} and Gaia \citep{gaia} will give extremely high accuracies in velocities and positions, LAMOST \citep{lamost} and part of the SDSS-III project, APOGEE and SEGUE-2 \citep{rockosi09}, will give high-quality spectra and therefore stellar parameters and chemical abundances of millions of stars over all the sky. These more accurate properties and larger samples will increase our knowledge of the Milky Way and answer wider questions about the formation and evolution of spiral galaxies.\\
As data sets grow it becomes of prime importance to create efficient and automatic tools capable of producing robust results in a timely manner. A standard technique for estimating parameters from data is the maximum-likelihood method. In a survey such as SEGUE, which contains more than 240,000 stellar spectra, each with more than 3000 flux measurements, it becomes extremely time-consuming to do spectral fitting with a brute-force search on the multi-dimensional likelihood surface, even more so if one wants to explore the errors on the recovered parameters.\\
Efforts to reduce the computational burden of characterizing the likelihood surface include Markov-Chain Monte-Carlo methods \citep{monte_carlo}, where a chain is allowed to explore the likelihood surface  to determine the global solution. An alternative method is to start at an initial estimate and hope that the likelihood surface is smooth enough that a gradient search will converge on the solution. An example of this approach can be found in \citet{ap06} and \citet{gray01}, based on the Nelder-Mead downhill simplex method \citep{simplex}, where the derivatives of the likelihood function give the direction toward the maximum. In both cases, the likelihood is evaluated only partially, leaving large amounts of parameter space untouched. Moreover, the time needed to find the maximum depends on the starting point and the steps used to evaluate the next point.  

Another parameter estimation method utilizes neural networks, for example \citet{snider} and \citet {re_fiorentin07}. While these non-linear regression models can obtain quick accurate results, they are entirely dependent on the quality of the training data, in which many parameters must be previously known. Careful attention must also be paid to the sampling of the model grid which is used to generate the neural network.

An alternative approach to spectral analysis is the Massively Optimized Parameter Estimation and Data compression method \citep[\small MOPED \normalsize \footnote{\small MOPED \normalsize is protected by US Patent 6,433,710, owned by The University Court of the University of Edinburgh (GB)},][]{h00}. This novel approach to the maximum likelihood problem involves compressing both data and models to allow very rapid evaluation of a set of parameters. The evaluation is fast enough to do a complete search of the parameter space on a finely resolved grid of parameters. Using carefully constructed linear combinations, the data are weighted and the size is reduced from a full spectrum to only one number per parameter. This number, with certain caveats (discussed later), contains all the information about the parameter contained in the full data. The method has been successfully applied in the fields of CMB analysis \citep{bond_cmb}, medical image registration \footnote{http://www.blackfordanalysis.com} and galaxy shape fitting \citep{vespa}. A complete background of the development of the \small MOPED \normalsize algorithm can be found in \citet[hereafter T97]{t97}, \citet[hereafter H00]{h00} and \citet{panter03}.

We present a new derivative of \moped, \Max~ (MAssive compression of $\chi^2$), to analyze stellar spectra.  To estimate the metallicity history of a galaxy for instance,  \moped~  needs models where the spectra are the sum of single stellar populations. Here we study one  population, meaning  the metallicity has to be estimated from the spectrum of each single star. It is therefore necessary to develop a specific tool for this task:   MA$\chi$. In Sect.~\ref{method} we describe the method, giving a summary of the derivation of the algorithm. We then test the method in Sect.~\ref{segue} on the basic stellar atmosphere  parameter estimation of a sample of low-resolution stars of SEGUE, where we compare our results with the SEGUE Stellar Parameter Pipeline in Sect.~\ref{comparison_segue} and finally check the method for a small sample of high-resolution spectra in Sect.~\ref{hires}, where we compare our results with a more classical approach for a small sample of VLT-UVES spectra. A summary of our conclusions is given in Sect.~\ref{fin}.

%%%%%%%%%%%%%%%%%%%%%%%%%%%%%%%%%%%%%%%%%%%%%%%%%%%%%%%%%%%%%%%%%%%%%%

\section{Method}\label{method}
In this section we describe our method to treat the likelihood surface. We first review the classical maximum-likelihood method, where a parametric model is used to describe a set of data.Then we present the algorithm that compresses the data and hence speeds the likelihood estimation. Finally we show the proof for the lossless nature of the compression procedure.

\subsection{Maximum-likelihood description}

Suppose the data (e.g. the flux of a spectrum) are represented by $N$ real numbers $x_1, x_2, ..., x_N$, which are arranged in an $N-$dimensional vector $\mathbf{X}$. They are the flux measurements at $N$ wavelength points. Assume that each data point $x_i$ has a signal part $\mu_i$  and a noise contribution $\sigma_i$
\begin{equation}
x_i = \mu_i + \sigma_i.
\end{equation}
If the noise has zero mean, $\langle \mathbf{X} \rangle = \langle \boldsymbol{\mu} \rangle$, we can think of $\mathbf{X}$ as a random variable of a probability distribution $L(\mathbf{X}, \mathbf{\Theta})$, which depends in a known way on a vector $\mathbf{\Theta}$ of $m$ model parameters. In the present case $m=3$ with
\begin{equation}\label{theta}
\mathbf{\Theta} = (\theta_1,\theta_2, \theta_3) = (\mathrm{[Fe/H],T_{eff}},\log g),
\end{equation}
with the iron abundance $\mathrm{[Fe/H]}$, effective temperature $\mathrm{T_{eff}}$ and surface gravity $\log g$. A certain combination of the single parameters $\theta_j$ ($j = 1,2,3$) also produces a  theoretical model $\mathbf{F} = \mathbf{F} (\mathbf{\Theta})$, and we can build an $m$-dimensional grid of theoretical models by varying the $\theta_j$ parameters. If the noise is Gaussian and the parameters have uniform priors, then the likelihood
\begin{equation}
L_f(\mathbf{X,\Theta}) = \frac{1}{(2 \pi)^{N/2}\sqrt{\sigma_{\mathrm{av}}^2}} \exp{\frac{-\chi^2}{2}}
\end{equation}
gives the probability for the parameters, where $\sigma_{\mathrm{av}}^2$ is the averaged square of the noise of each data point and
\begin{equation}
\label{chi_full}
\chi^2 = \sum_{i=0}^N \frac{(F_i-\mu_i)^2}{\sigma_i^2}.
\end{equation}
The position of its maximum estimates the set of parameters $\mathbf{\Theta_0}$ that best describe the data \textbf{X}. The fit between $\mathbf{F}(\mathbf{\Theta_0})$ and $\mathbf{X}$ is good if $\chi^2 < \mathrm{DOF}$, where $\mathrm{DOF} = N - m$ are the degrees of freedom. In the most basic form one finds the maximum in the likelihood surface by exploring all points in the parameter space, with each likelihood estimation calculated using all $N$ data points. This procedure is of course very time-consuming if $N$ and $m$ are large.

\subsection{Data compression}

In practice not all data points give information about the parameters, because either they are noisy or not sensitive to the parameter under study. The \moped  algorithm uses this knowledge to construct weighting vectors which neglect some data without losing information. A way to do this is by forming linear combinations of the data.

Let us  compress the information of a given parameter $\theta_j$ from our spectrum. The idea is to capture as much information as possible about this particular parameter. We define our weighting vector $\mathbf{b}_j = (b_{j,1}, b_{j,2},...,b_{j,N})$ as 
\begin{equation}\label{eq:b}
\mathbf{b}_{j} = \frac{1}{|\mathbf{b}_{j}|} \frac{1}{\sigma^2} \frac{\partial \mathbf{F}}{\partial \theta_{j}},
\end{equation}
where $|\mathbf{b}_{j}| = \sqrt{\mathbf{b}_j \cdot \mathbf{b}_j}$. This definition assures that each data point of the spectrum is less weighted  if $\sigma$ is large and more weighted if the sensitivity in the flux -- meaning the derivative of $\mathbf{F}$ respect to the parameter $\theta_j$ -- is high. To create this vector we need, on one hand, the information about the behavior of the parameters from a theoretical fiducial model, and  on the other hand, the error measurement from the data. 

\begin{figure}
\centering
\includegraphics[scale=0.5,angle=90]{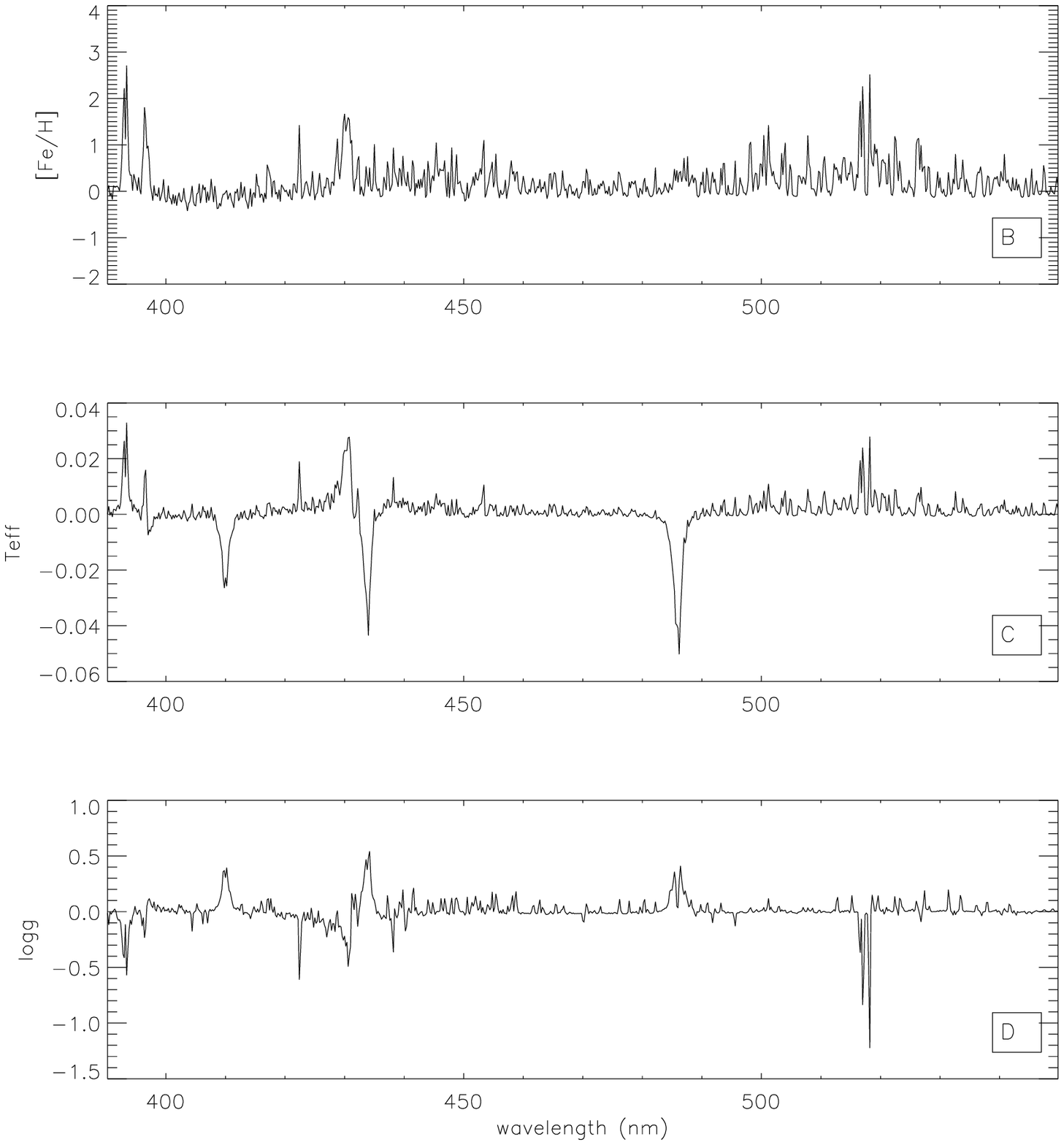}
\caption{ (A) Fiducial model with parameters $\mathrm{[Fe/H]} = -2.0$, $\mathrm{T_{eff}} = 6000 $ K and $\log g = 4.0$.  Other panels indicate the weighting vectors according to Eq.~(\ref{eq:b}) for metallicity (B), temperature (C) and surface gravity (D). }
\label{weights}
\end{figure}

Figure~\ref{weights} shows the example of the weighting vectors for the parameters of the set (\ref{theta}) using a fiducial model with $\mathrm{[Fe/H] = -2.0}$, $\mathrm{T_{eff}} = 6000  $ K and $\log g = 4.0$. The upper panel corresponds to the synthetic spectrum at the SDSS resolving power (R = 2000) and the panels B, C and D to the weighting vector for metallicity, effective temperature and surface gravity, respectively. This figure is a graphical representation of Eq.~(\ref{eq:b}), and helps to  better visualize the weight in relevant regions of the spectrum.

For metallicity (panel B), the major weight is concentrated in the two peaks at 3933 and  3962 \AA, corresponding to the CaII K and H lines seen in the synthetic spectrum in panel A. The second important region with weights is close to 5180 \AA, corresponding to the MgIb triplet. A minor peak is seen at 4300 \AA, which is the G-band of the CH molecule. These three features are metallic lines, therefore show a larger dependency on metallicity than the rest of the vector, which is dominated by a zero main weight.

For the temperature (panel C) the greatest weight is focused on the hydrogen Balmer lines at 4101, 4340 and  4861 \AA. Peaks at the previous metallic lines are also seen, but with a minor amplitude.  Finally, the surface gravity (panel D) presents the strongest dependence on the \mgi~ triplet. The wings of the \caii and Balmer lines also present small dependence.  As in the case for metallicity, the rest of the continuum is weighted by a mean close to zero.

The weighting is done by multiplying these vectors with the spectrum, where the information kept from the spectrum is given by the peaks of the weighting vector.  We define this procedure as ``compression'', where the information about the parameter $\theta_j$ is expressed as
\begin{equation}\label{y}
y_{j,\mathbf{X}} = \mathbf{b}_j \cdot \mathbf{X}, \indent \indent  j= 1,2,3
\end{equation}
with $ \mathbf{X}$ the spectrum. With the weighting vector defined in Eq.~(\ref{eq:b}) we can compress data ($\mathbf{X}$) and model ($\mathbf{F}$) by using Eq.~(\ref{y}). Because $\mathbf{b}_j$ and $\mathbf{X}$ (or $\mathbf{F}$) are $N-$dimensional, the product $y_j$ in (\ref{y}) is a number, which stores the information about the parameter $\theta_j$.

 In order to perform a compression for all the parameters simultaneously, we require that $y_k$ is uncorrelated with $y_j$, with $j \neq k$. This means that the \textbf{b}-vectors must be orthogonal, i.e. $\mathbf{b}_k \cdot \mathbf{b}_j = 0$. Following the procedure of H00 we find the other  $y_k = \mathbf{b}_k \cdot \mathbf{X}$ by the Gram-Schmidt ortogonalization

\begin{equation}
\mathbf{b_k} = \frac{1}{|\mathbf{b}_{k}| }\left[\frac{1}{\sigma^2}\frac{\partial \mathbf{F}}{\partial \theta_k} - \sum_{q =1}^{m-1}\left( \frac{\partial \mathbf{F}}{\partial \theta_k} \mathbf{b}_q\right) \mathbf{b}_q \right].
\end{equation}
For $m = 3$  we have the numbers  $y_{1,\mathbf{X}}$,  $y_{2,\mathbf{X}}$, $y_{3,\mathbf{X}}$, for the data and $y_{1,\mathbf{F}}$,$y_{2,\mathbf{F}}$,$y_{3,\mathbf{F}}$ for the models, corresponding to the parameters of $\mathbf{\Theta}$ given in Eq.~(\ref{theta})

\begin{figure}
\centering
\includegraphics[scale=0.3,angle=90]{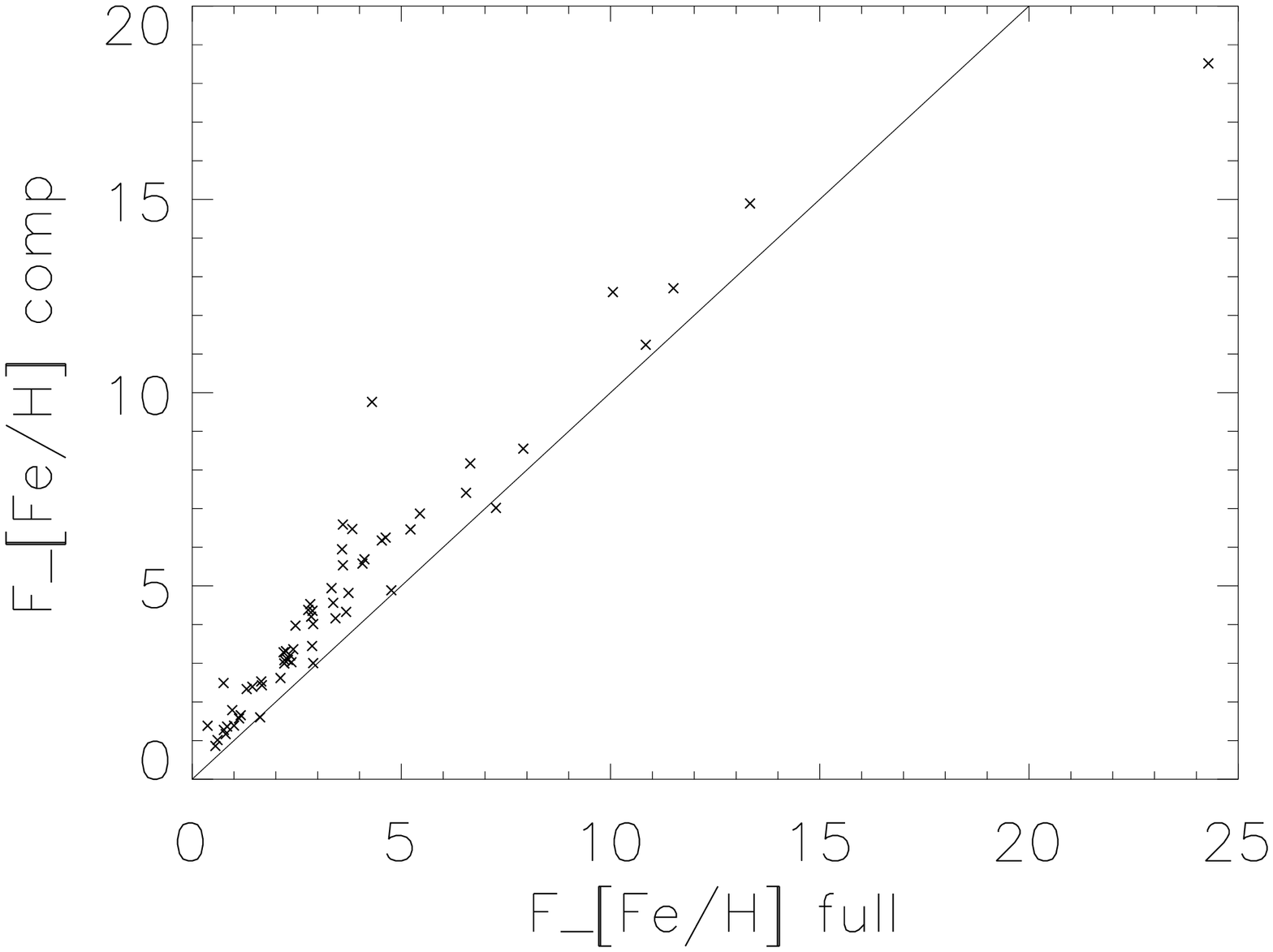}
\includegraphics[scale=0.3,angle=90]{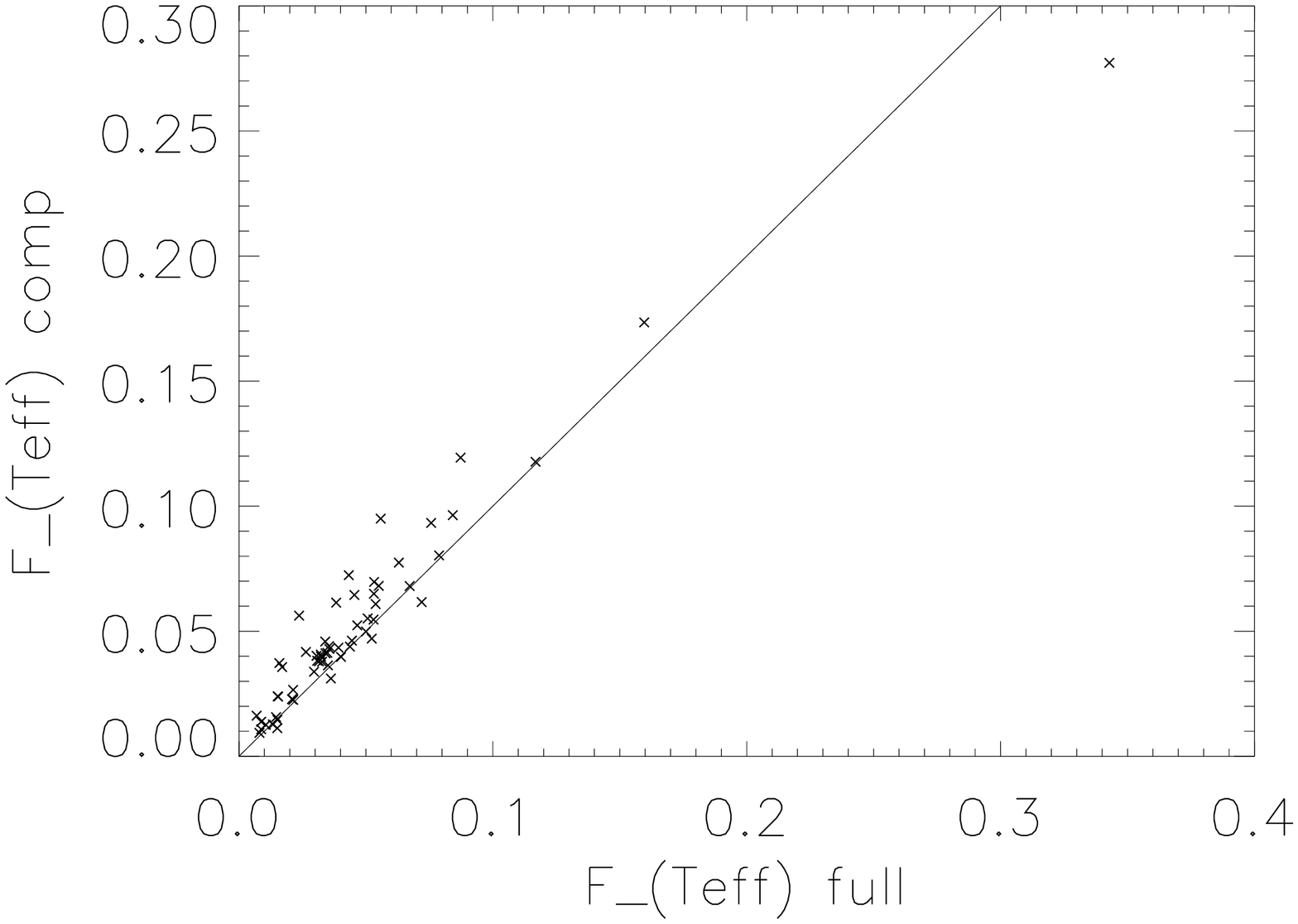}
\caption{Correlations between the Fisher matrix values obtained from the full and compressed likelihoods for metallicity (upper panel) and effective temperature (lower panel) for 75 randomly selected stars of SEGUE. The line corresponds to the one-to-one relation.}
\label{fisher}
\end{figure}

\subsubsection*{Goodness of fit}

The definition of $\chi^2$, which is the sum of the differences of the data and the model, motivates us to define our goodness of fit. The ``compressed'' $\chi^2_\mathrm{c}$ is the sum of the differences of the compressed data and model
\begin{equation}
\chi_{\mathrm{c,\theta_j}}^2 = \frac{1}{\sigma_{\mathrm{av}}^2} (y_{j,\mathbf{X}} - y_{j,\mathbf{F}})^2,
\end{equation}
which gives the probability for the parameter $\theta_j$.
\begin{equation}
L_{c,j}(\mathbf{X},\theta_j) =  \frac{1} {(2 \pi)^{N/2}  \sqrt{\sigma_{\mathrm{av}}^2}} \exp{ \frac{-\chi_{\mathrm{c},\theta_j}^2}{2}}
\end{equation}
Because the $y_j$ numbers are by construction uncorrelated, the ``compressed" likelihood of the parameters is obtained by multiplication of the likelihood of each single parameter $L_{c,j}(\mathbf{X},\theta_j)$
\begin{equation}
L_c(\mathbf{X},\mathbf{\Theta}) =\prod_{j=1}^m L_{c,j}(\mathbf{X},\theta_j)= \frac{1} {\left( (2 \pi)^{N/2}  \sqrt{\sigma_{\mathrm{av}}^2}\right)^m} \exp{\frac{-\chi_{\mathrm{c}}^2}{2}},
\end{equation}
where
\begin{equation}\label{chi_c}
\chi_{c}^2 = \frac{1}{\sigma_{\mathrm{av}}^2} \sum_{j = 0}^m (y_{j,\mathbf{X}} - y_{j,F})^2
\end{equation}
is the compressed $\chi^2$. 

As in the classical approach, the peak of the compressed likelihood estimates the parameters $\mathbf{\Theta}_0$ that generate the model $\mathbf{F}(\mathbf{\Theta}_0)$, which reproduces best the data $\mathbf{X}$.
The advantage of using the compressed likelihood is that it is fast, because the calculation of $\chi_c^2$ needs only $m$ iterations and not $N$ as in the usual $\chi^2$ computation.  The search of the maximum point in the likelihood is therefore done in an $m-$dimensional grid of $y-$numbers and not of $N-$length fluxes.% Here the power of the method: we deal with the number of parameters and not the number of datapoints.

\subsection{No loss of information by the compression}

The Fisher Information Matrix describes the behavior of the likelihood close to the maximum. Here we use it only to show the lossless compression offered by the \small MOPED \normalsize method, but a more extensive study is given in T97 and H00. \\
To understand how well the compressed data constrain a solution we consider the behavior of the logarithm of the likelihood $L$ near the peak. Bellow we denote a generic $L$ that can correspond to the compress likelihood $L_c$ or the full one $L_f$, because the procedure is the same. In the Taylor expansion the first derivatives $\partial (\ln L)/ \partial \theta_j$ vanish at the peak and the behavior around it is dominated by the quadratic terms
\begin{equation}\label{taylor}
\ln L =  \ln L_{\mathrm{peak}}+ \frac{1}{2} \frac{\partial^2 \ln L}{\partial \theta_{j} \partial \theta_{k}}\Big|_{\mathrm{peak}} \Delta \theta_{j} \Delta \theta_{k},
\end{equation}
The likelihood function is approximately a Gaussian near the peak and the second derivatives, which are the components of the inverse of the Fisher Matrix $\mathcal{F}_{jk}$, measure the curvature of the peak. Because the dependence of the parameters is not correlated, the matrix is diagonal with values
\begin{equation}
\mathcal{F}_{jj} = - \frac{\partial^2 \ln L}{\partial \theta_j^2}.
\end{equation}

\begin{figure}
\centering

\includegraphics[scale = 0.8]{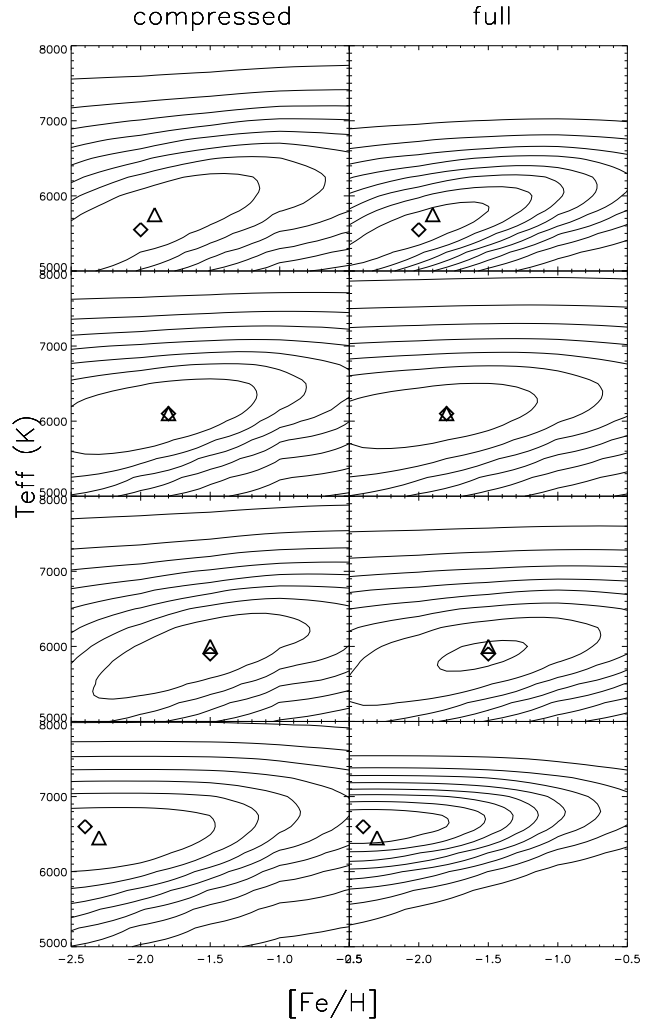}
\caption{Likelihoods of the compress data set (left) and full data set (right) in the parameter space of effective temperature and metallicity for four randomly selected SEGUE stars. In each panel eight equally spaced contour levels are plotted.  The triangle and diamond correspond to the maximum point of the compressed and the full data set, respectively. Both maxima lie in the first confidence contour level, meaning that both data sets reach the same solution for these two parameters.}
 \label{likes}
\end{figure}

In Fig.~\ref{fisher} we  plotted the values $\mathcal{F}_{\mathrm{[Fe/H]}}$ and $\mathcal{F}_{\mathrm{T}_{\mathrm{eff}}}$, corresponding to the parameters of metallicity and effective temperature, respectively, for likelihoods of 75 randomly selected stars of SEGUE, using the classical (full) and the compressed (comp) dataset. The line corresponds to a one-to-one relation. The values correlate well, as expected.  \\
Figure~\ref{likes} shows full and compressed likelihood surfaces of metallicity and temperature for four randomly selected stars, where eight equally spaced contour levels have been plotted in each case.  This is another way to visualize the statement that the Fisher Matrices correlate.  The curvature of both likelihoods is the same close to the peak, as predicted. In Fig.~\ref{likes} we  also over-plotted the maximum point of the compressed likelihood as a triangle and the maximum point of the full one as a diamond. It can be seen that the maximum points of the compressed and full data set lie inside the first contour level of the likelihoods. This shows that the compressed data set gives the same solution as the full one - even after the dramatic compression.

In the bellow parts the numerical values of $\chi^2$ correspond to the  classical reduced $\chi^2$.\\

%%%%%%%%%%%%%%%%%%%%%%%%%%%%%%%%%%%%%%%%%%%%%%%%%%%%%%%%%%%%%%%%%%%%%%%%%%%%

\section{Implementation on low-resolution spectra}\label{segue}

The sample of stars used to test the method are F-G dwarf stars. Based on the metallicity given by the SEGUE Stellar Parameter Pipeline, we chose metal-poor stars  as will be explained below. This choice allows us to avoid the problem of the saturation of metallic lines such as \caii K \citep{beers99}, which is a very strong spectral feature  that serves as a metallicity indicator in the low-resolution spectra from SEGUE. These stars fall in the temperature range where Balmer lines are sensitive to temperature and the spectral lines are not affected by molecules \citep{gray_book}. With these considerations it is correct to assume that the spectra will behave similar under changes of metallicity, temperature and gravity. This allows the choice of a random fiducial model from our grid of models for the creation of the weighting vectors, which will represent well the dependence of the parameters in all the stars.

\subsection{Data}
We  used a sample of SEGUE stellar spectra \citep{Yanny09}, part of the Seventh Data Release \citep[DR7,][]{dr7} of SDSS.  The survey was performed with the $2.5$-meter telescope at the Apache Point Observatory in southern New Mexico and contains spectra of $\sim 240,000$ stars. All the spectra have $ugriz$ photometry \citep{fukugita}.  From a color-color diagram we identified the dwarf stars and selected them using the sqlLoader \citep{sql} under the  constraints that the object must be a star and have  colors of $0.65 < u-g < 1.15$ and $0.05 <g-r< 0.55$. We verified the spectral type by  also downloading the values 'seguetargetclass' and 'hammerstype', which classify our stars mainly as F-G dwarfs.  We also constrained the metallicity to be in the range of $-999 < \mathrm{[Fe/H]} < -0.5$ . The latter values were taken from the new SEGUE Stellar Parameter Pipeline \citep{lee08_1, lee08_2, SSPPIII}, were the -999 indicates that the pipeline has not estimated the metallicity of this particular star. \\
Our final sample contains spectra of 17,274 stars with a resolving power of $R  \simeq 2000$ and a signal-to-noise up to 10. The wavelength range is  $3800-9200 \AA$ and the spectra are with absolute flux.

\subsection{Grid of models}\label{grid_sdss}
The synthetic spectra were created with the synthesis code SPECTRUM \citep{spectrum}, which uses the new fully blanketed stellar atmosphere models of \citet{Kurucz_models} and computes the emergent stellar spectrum under the assumption of local thermal equilibrium (LTE). The stellar atmosphere models assume the solar abundances of \citet{gs98} and a plane parallel line-blanketed model structure in one dimension.  For the creation of the synthetic spectra, we set a microturbulence velocity of 2 km/s, based on the atmosphere model value. The line-list file and atomic data were taken directly from the SPECTRUM webpage \footnote{http://www1.appstate.edu/dept/physics/spectrum/spectrum.html}. In these files, the lines were taken from the NIST Atomic Spectra Database \footnote{http://physics.nist.gov/PhysRefData/ASD/index.html} and the Kurucz webside \footnote{http://kurucz.harvard.edu/linelists.html}. No molecular opacity was considered in the model generation. 

We created an initial three dimensional grid of synthetic spectra starting from the ATLAS9 Grid of stellar atmosphere models of  \citet{atlas9} by varying the parameters $\mathbf{\Theta}$ of Eq.~(\ref{theta}). They cover a wavelength range from 3800 to 7000 $\AA$ in steps of $\Delta  \lambda = 1 \AA$, based on the wavelength range of SDSS spectra together with our line-list. This wavelength range is broad enough for a proper continuum subtraction, as described in Sect.~\ref{preparation}. The spectra  have an absolute flux and were finally smoothed to a resolving power of $R = 2000$, according to SDSS resolution.

In order to have a finer grid of models, we  linearly interpolated the fluxes created for the initial grid. It has models with $-2.5 \leq \mathrm{[Fe/H]} \leq -0.5$ with  $\Delta \mathrm{[Fe/H]} = 0.1$ dex, $5000 \leq \mathrm{T_{eff}} \leq 8000$ with $\Delta  \mathrm{T_{eff}} = 50$ K and $3.5 \leq \log g \leq 5$ with $\Delta \log g = 0.1$ dex. The scaling factor for the normalization varies linearly from 0.85 to 1.15 in steps of 0.01 (see below). Our final grid has $21 \times 61 \times 16 \times 30 = 614,880$ models. We set a linearly varying $\mathrm{[\alpha/Fe]}$ of $\mathrm{[\alpha/Fe]} = [0.2,0.4]$ for stars in the metallicity range of  $-0.5 \geq \mathrm{[Fe/H] \geq -1.0}$ and $\mathrm{[\alpha/Fe]} = 0.4$ for stars with $\mathrm{[Fe/H] \le -1.0}$, to reproduce the abundance of $\alpha$ elements in the Milky Way, as in \citet{lee08_1}. These varying $[\alpha/\mathrm{Fe}]$ abundances were  also calculated with interpolation of fluxes created from solar and $\alpha-$ enhanced ([$\alpha$/Fe] = 0.4) stellar atmosphere models. We did not include the $\alpha$ abundances as a free parameter, because we aim to estimate the metallicity from lines of $\alpha$ elements (Ca, Mg), meaning we would obtain a degeneracy in both parameters. We prefered to create a grid where the $\alpha$ abundances were already 'known'.

 \begin{figure}
\centering
\includegraphics[scale=0.8, angle=90]{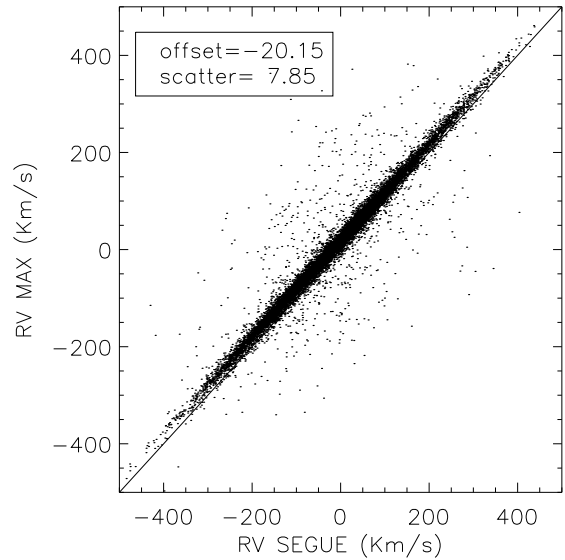}
\caption{Radial velocities found using minimum fluxes of strong lines given by Table~\ref{tab_lines} (MAX) compared with those of the SEGUE database. The difference of the radial velocities given by the SEGUE database from those obtained by our method is indicated in the legend as offset, with its standard deviation as scatter.}
\label{fig:rv}
\end{figure}

The grid steps between the parameters are smaller than the accuracies expected from the low resolution SEGUE data. The extra time required to calculate the compressed likelihood in this finer grid is not significant, and retaining the larger size allows us to demonstrate the suitability of the method for future more accurate data capable of using the full grid.

\begin{table}
\caption{Strongest lines in F-G dwarf stars. The first column indicates the name of the line and the second its wavelength position in \AA.}
\label{tab_lines}
\centering
\begin{tabular}{c c}
\hline\hline
 line & $\lambda$ \AA  \\
\hline
 Ca $II K$ & 3933 \\
Ca $II H$ & 3962 \\
H$\delta$ & 4101 \\
H$\gamma$ & 4340 \\
H$\beta$ & 4861\\
Mg b  triplet & 5183\\
\hline
\end{tabular}
\end{table}

\subsection{Matching models to data}\label{preparation}

Data and model needed to be  prepared for the analysis.  First of all, the data needed to be corrected from the vacuum wavelength frame of the observations to the air wavelength frame of the laboratory. Secondly, we needed to correct for the Doppler effect $zc$. This was done by using the flux minimum of the lines indicated in Table~\ref{tab_lines} except for the \mgi b triplet, because this feature is not seen clearly enough in every spectrum, driving in our automatic $zc$ calculation to unrealistic values. A comparison of the radial velocity found with this method with the value given by the SEGUE database f is shown in Fig.~\ref{fig:rv}. The difference of the radial velocities (SEGUE - MAX) has a mean (offset) of -20.15 km/s  with standard deviation (scatter) of 7.85 km/s. The effect of this difference on the final parameter estimation is discussed in  Sect.\ref{comparison_segue}. Once our data were corrected for the Doppler effect, they had to be interpolated to the wavelength points of the models to get the same data points.  \\
For automatic fitting of an extensive sample of stars with a large grid of models, we decided to normalize the flux to a fitted continuum. In this way the dependence of the parameters is this is not past, it is a fact that happens now too) concentrated mainly on the line profiles. The choice of the normalization method is a difficult task, because none of them is perfect. It becomes especially difficult in regions where the spectrum shows too many strong absorption features, as the \caii lines of Table~\ref{tab_lines}. Using the same method for models and data, difficult regions behave similar in both cases, which allowed us a final fitting in these complicated regions as well. We adopted the normalization method of \citet{ap06}, because it works well for the extended  spectral range of SDSS spectra. It  is based on an iteratively polynomial fitting of the pseudo-continuum, considering only the points that lie inside the range of $4 \sigma$ above and $0.5 \sigma$ below. Then we divided the absolute flux by the final pseudo-continumm. Due to noise, the continuum of the final flux may not necessary be at unity. We included a new degree of freedom in the analysis that scaled this subtracted continuum.  This scaling factor for the normalization is then another free parameter, which means that we have finally four parameters to estimate -- three stellar atmosphere parameters and the scaling factor. The final step is to choose the spectral range for the analysis.

 \begin{figure}
\centering
\includegraphics[width=6.5cm, angle=90]{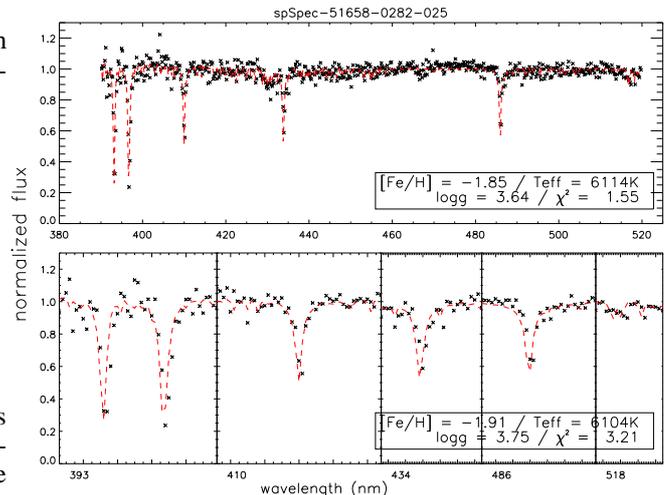}
\caption{Example of the fit between a randomly selected SEGUE star (crosses) and a synthetic spectrum (red dashed line). The legend indicates the stellar atmosphere parameters of the model and the value of the reduced $\chi^2$ of the fit. The upper panel is the fit using all the points of the spectral region [3850,5200] {\AA}. The lower panel is the fit using only the data points where the lines of Table~\ref{tab_lines} are located.}
\label{fit}
\end{figure}

\subsection{Compression procedure}\label{compression}

We chose the fiducial model with parameters $\mathrm{[Fe/H]} = -2.0, \mathrm{T_{eff}}=6000$ K, $\log g = 4.0 $ and $\mathrm{a}_s = 1$, to calculate the weighting vectors, \textbf{b}$_{\mathrm{[Fe/H]}}$, \textbf{b}$_{\mathrm{T_{eff}}}$, \textbf{b}$_{\log g}$ and \textbf{b}$_{\mathrm{a}_s}$, corresponding to metallicity, effective temperature, gravity and scaling factor for normalization, respectively.  Then, we calculated the set of $y-$numbers using Eq.~(\ref{y}) for each point in the grid of synthetic spectra by projecting the \textbf{b}-vectors onto the spectrum, resulting in a four dimensional $y-$grid, with every point a single number.  With the respective \textbf{b}-vectors we calculated the $y-$numbers for the observed spectra, the expression (\ref{chi_c}) for every point in the grid and finally we found the minimum value in the grid which corresponds to the maximum point of the compressed likelihood. 

To refine our solution we found the ``real'' minimum $\chi^2$ with a quadratic interpolation.  For the errors, we looked at  the models within the confidence contour of $\chi^2 = \chi^2_{min} + \Delta \chi^2$ with $\Delta \chi^2 = 4.72$  corresponding to the  $1 \sigma$ error in a likelihood with four free parameters \citep[for further details see chapter 14 of][]{num_rec}.

\begin{figure*}
\centering
\includegraphics[width=9cm,angle=90]{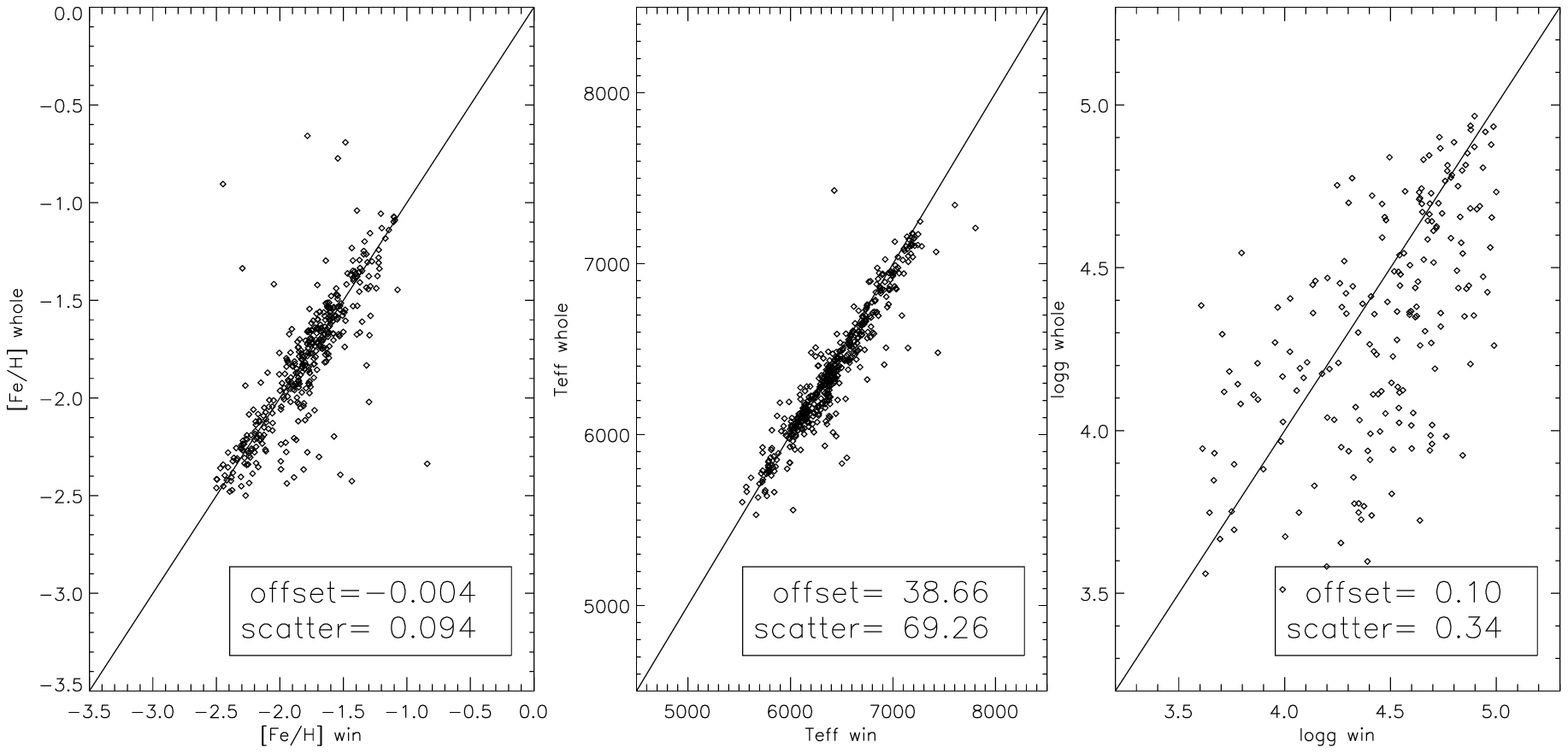}
\includegraphics[width=9cm,angle=90]{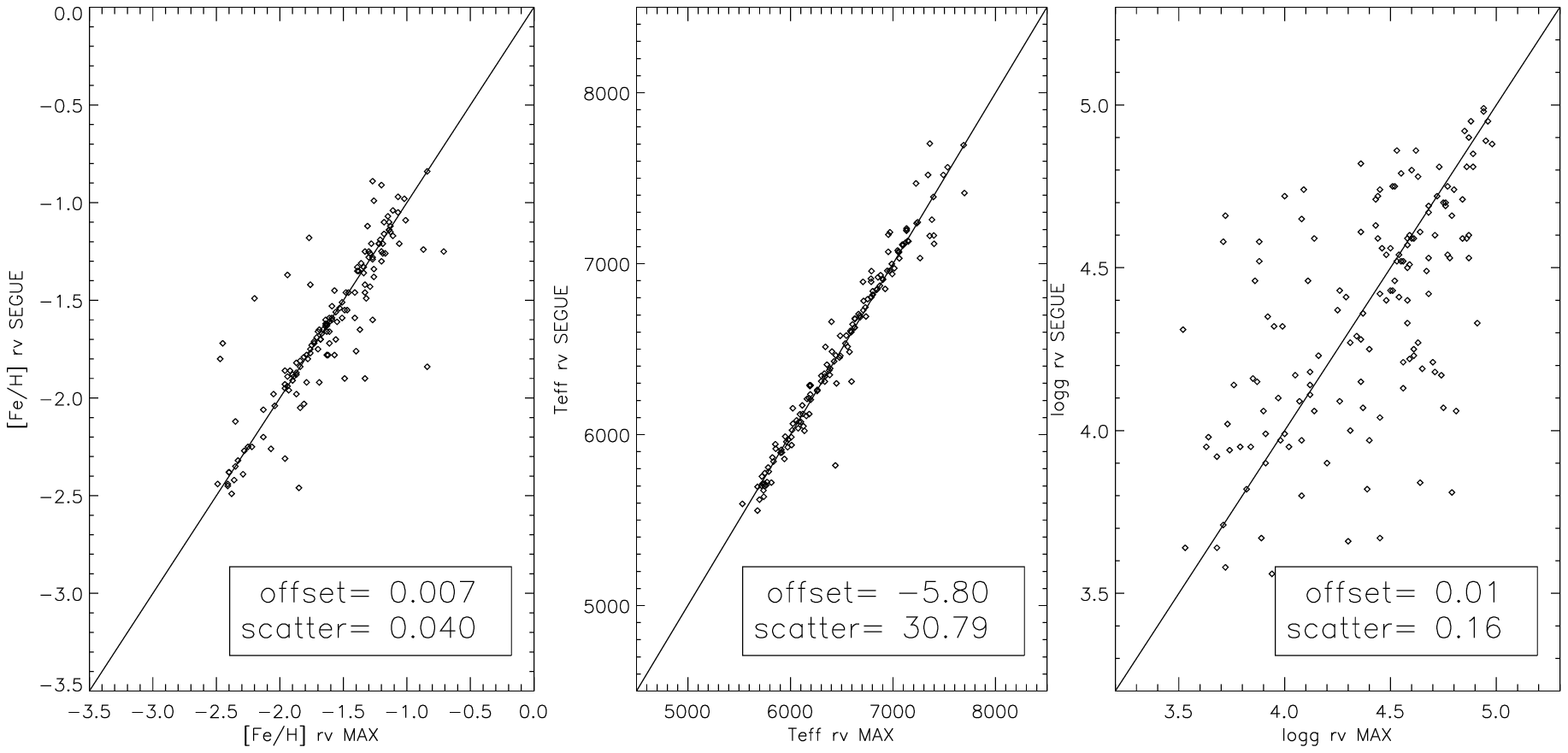}
\caption{Upper panel: Metallicity (left), effective temperature (middle) and surface gravity (right) obtained using MA$\chi$ for the entire spectral range of [3900, 5200] $\AA$ (whole) compared to that with selected spectral windows (win) for a sample of 150 randomly selected stars.  The offset (mean difference of ``win - whole'') of the results and its scatter (standard deviation)  is indicated at the bottom right of each plot. The line has a slope of unity. Lower panel: as upper panel, but investigating the influence of using our radial velocities (rv MAX) or the SEGUE ones (rv SEGUE). Offset and scatters are calculated from the difference  ``SEGUE-MAX''.}
\label{corr_intrinsic}
\end{figure*}

\section{Application to SEGUE spectra}\label{comparison_segue}

We chose to use the wavelength range of  [3850,5200]~\AA  as information about metallicity, temperature and gravity, which is available in the lines listed in Table~\ref{tab_lines}.  These spectral lines are strong in F-G stars, meaning they can be identified at low resolution without difficulty. The wings of  Balmer lines are sensitive to temperature and the wings of strong Mg lines are sensitive to gravity \citep{fuhrmann98}. Beacuse iron lines are not strong enough to be distinguished from the noise of our spectra,  the \caii K and \mgi b lines are our indicators of metallicity \citep{beers00,ap06,lee08_1}.

 It is important to discuss the carbon feature known as the G-band at 4304 \AA~ in our spectral window.  The fraction of carbon-enhanced metal-poor (CEMP) stars is expected to increase with decreasing metallicity \citep{beers92} to about 20\% at below $\mathrm{[Fe/H]} = -2.0$ \citep{lucatello},possibly implying a strong carbon feature in the observed spectrum. As discussed by \citet{marsteller}, a strong G-band may also affect the measurement of the continuum at the \caii lines, which could result in an underestimation of the stellar metallicity.  We commented in Sect.~\ref{method} on the influence of the lines in the weighting vectors used for the compression and we saw that the G-band also displayes minor peaks (see Fig.~\ref{weights}). The role of this minor dependence compared with those from the \caii, Balmer and \mgi~ lines can be studied by comparing the final parameter estimation when using the whole spectral range or only those regions with the lines of Table~\ref{tab_lines}, where the G-band is not considered.  A further motivation for performing an analysis of an entire range against spectral windows is also explained below. The implications of this tests in terms of fraction of CEMP stars is discussed below.

\subsection{Whole domain vs. spectral windows}

The classical $\chi^2$ fitting procedure uses every datapoint; therefore, a straight-forward method to speed up the analysis would be to mask those parts of the spectrum which do not contain information about the parameters, essentially those that are merely continuum. Certainly, by considering only the \caii, Balmer and \mgi b lines, the number of operations becomes smaller, increasing the processing speed of $\chi^2$ calculations. This is clumsy however: an empirical decision must be made about the relevance of pixels, and no extra weighting is considered - and the time taken for the parameters estimation is still long if one decides to do it for many spectra. The use of the \textbf{b}-vectors in the \Max~ method means that very little weight is placed on pixels which do not significantly change with the parameter under study, automatically removing the sectors without lines. 

The remarkable result of the \Max~ method is that it is possible to determine the maximum of the compressed likelihood in 10 milliseconds with a present day standard desktop PC. This is at least 300 times faster than the same procedure when doing an efficient evaluation of the uncompressed data. \\
The 1 $\sigma$ confidence contours indicate errors in the parameters of  0.24 dex in metallicity on average, 130 K in temperature and 0.5 dex in gravity. Examples of fits are shown in Fig.~\ref{fit}. The upper plot shows the fit between a randomly selected SEGUE spectrum and the best model - the crosses correspond to the observed spectrum and the dashed red line to the model, with stellar atmosphere parameters in the legend, as well as the resulting reduced $\chi^2$ of the fit. The plot in the bottom is the fit of the same star, but considering only the data points within the line regions identified in Table~\ref{tab_lines}, as discussed above. Again, the red dashed line indicates the model with parameters in the legend. The $\chi^2$ of the fit is also given. 

Because the lines contain the most information about metallicity, temperature and gravity, we expect to obtain the same results whether we use all the data points or only those corresponding to the lines.  The legends in Fig.~\ref{fit} show  the parameters estimated in both analyses. The small differences between them are within the 1 $\sigma$ errors. We plotted in the upper panel of Fig.~\ref{corr_intrinsic} the comparison for metallicity (left panel), temperature (middle) and gravity (right) of our sample of  stars when using the whole spectral range (``whole'') or only spectral windows with the lines (``win''). The offsets and scatters of the distribution as well as the one-to-one relation are indicated in the legend of each plot. Here we randomly selected 150 stars to plot in Fig.~\ref{corr_intrinsic} to visualize better the correlations of the results. Offsets and scatters are calculated with the entire star sample. Metallicity has excellent agreements, with a small scatter of 0.094 dex, as shown in the legend of the plot. Temperature has a negligible offset of 38 K and a scatter of 69 K, which is also less than the accuracies obtained in the temperature estimation.  Gravity has the largest scatter offset of 0.1 dex and a scatter of 0.34 dex, but this is still within the errors.

The negligible offsets obtained when using the entire spectral range against the limited regions is an encouraging result in terms of the effect of the G-Band on the spectrum. As pointed out above, the dependence on this feature in the parameters under study is not as strong as the rest of the lines of Table~\ref{tab_lines}. This is translated into less weight for the compression, as seen in Fig.~\ref{weights}, meaning the G-band does not play a crucial role in our compressed data set for the parameter recovery. 

Let us remark that this does not mean a lack of CEMP stars in our sample,  we simply do not see them in the compressed space. A spectrum with a strong observed CH molecule gives a similar \underline{compressed} $\chi^2$ as one with a weak one, because the compression does not consider variations in carbon abundance. The \underline{full} $\chi^2$, on the other hand, will certainly be larger for the spectrum with a strong G-band, because our models do not have different carbon abundances.  

In any case, the fraction of CEMP stars is high and it is certainly interesting to locate them with the \Max~ method in the future. This implies that we need to create another dimension in the grid of synthetic spectra -- models with varying carbon abundances -- and increase the number of parameters to analyze to five. One more dimension certainly means a heavier grid of models, but in terms of parameter recovery, there is no big difference to use four or five $y-$numbers for the compressed $\chi^2$ calculation. This study goes beyond the scope of this paper, however, where we only introduce the \Max~ method.

\subsection{Effect on radial velocities}

As mentioned in Sect.~\ref{preparation} and seen in Fig.~\ref{fig:rv}, radial velocities given by the SEGUE database have a mean difference of 20 km/s compared to ours. To study the effect of this difference in the final parameter estimation, we  analyzed  a sub-sample of randomly selected 2,670 stars of our sample using SEGUE radial velocities. The comparison of the metallicity, effective temperature and surface gravity obtained when using our radial velocities (``rv MAX'') and the SEGUE ones (``rv SEGUE'') can be seen in the lower panel of Fig.~\ref{corr_intrinsic}. In this plot, also a small sample of randomly selected 150 stars was used  to  better visualize the results, but offsets and scatters were calculated considering the 2,670 stars. A difference in 20 km/s produces  offsets and scatters in the three parameters, with $0.007 \pm 0.040$ dex for metallicity, $-5.08  \pm 30.79$ K for temperature and $0.01 \pm 16$ dex for gravity. These offsets are negligible when compared with the $1 \sigma$  errors obtained the parameter estimation. 
\rm

\begin{figure*}
\centering
\includegraphics[scale=0.8,angle=90]{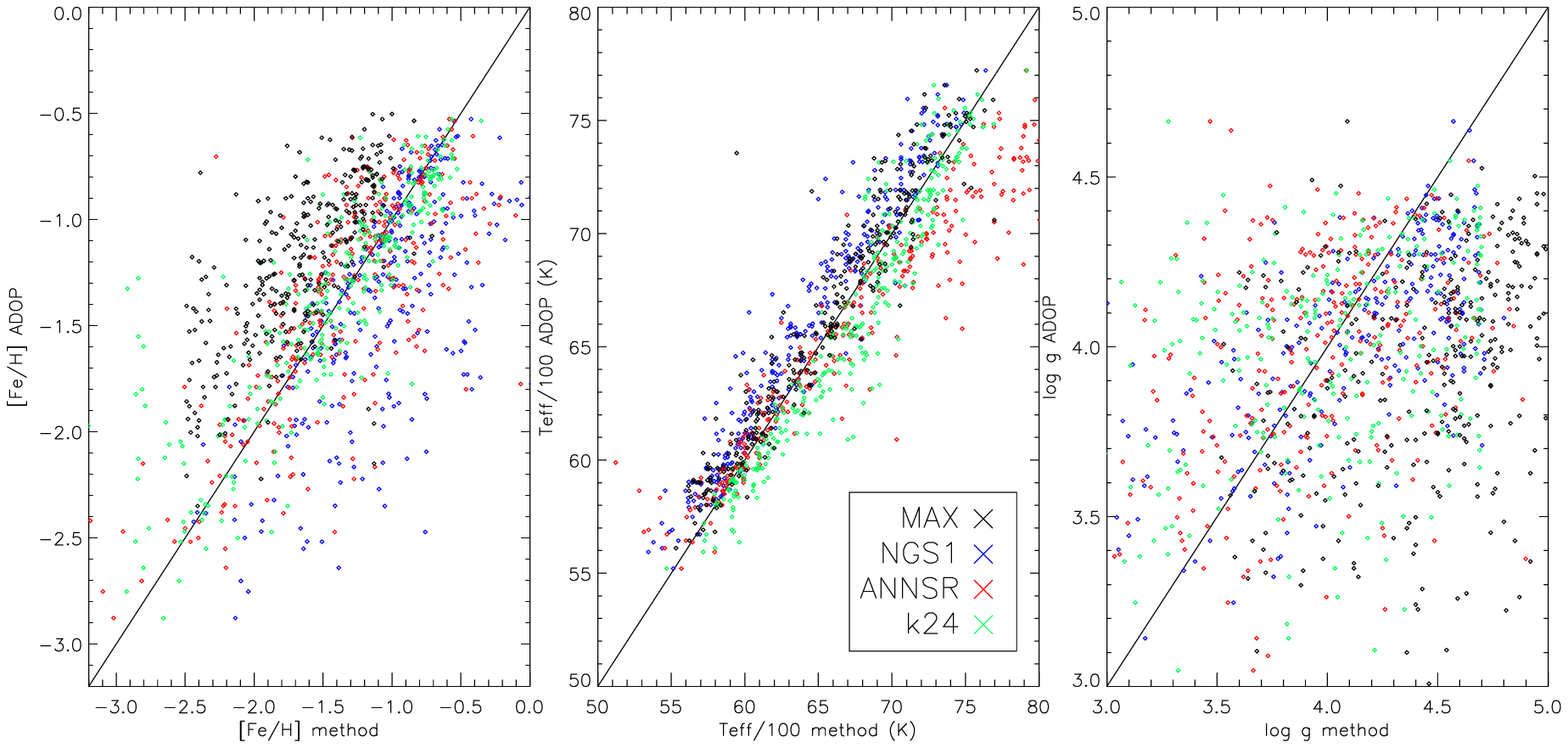}
\includegraphics[scale=0.8,angle=90]{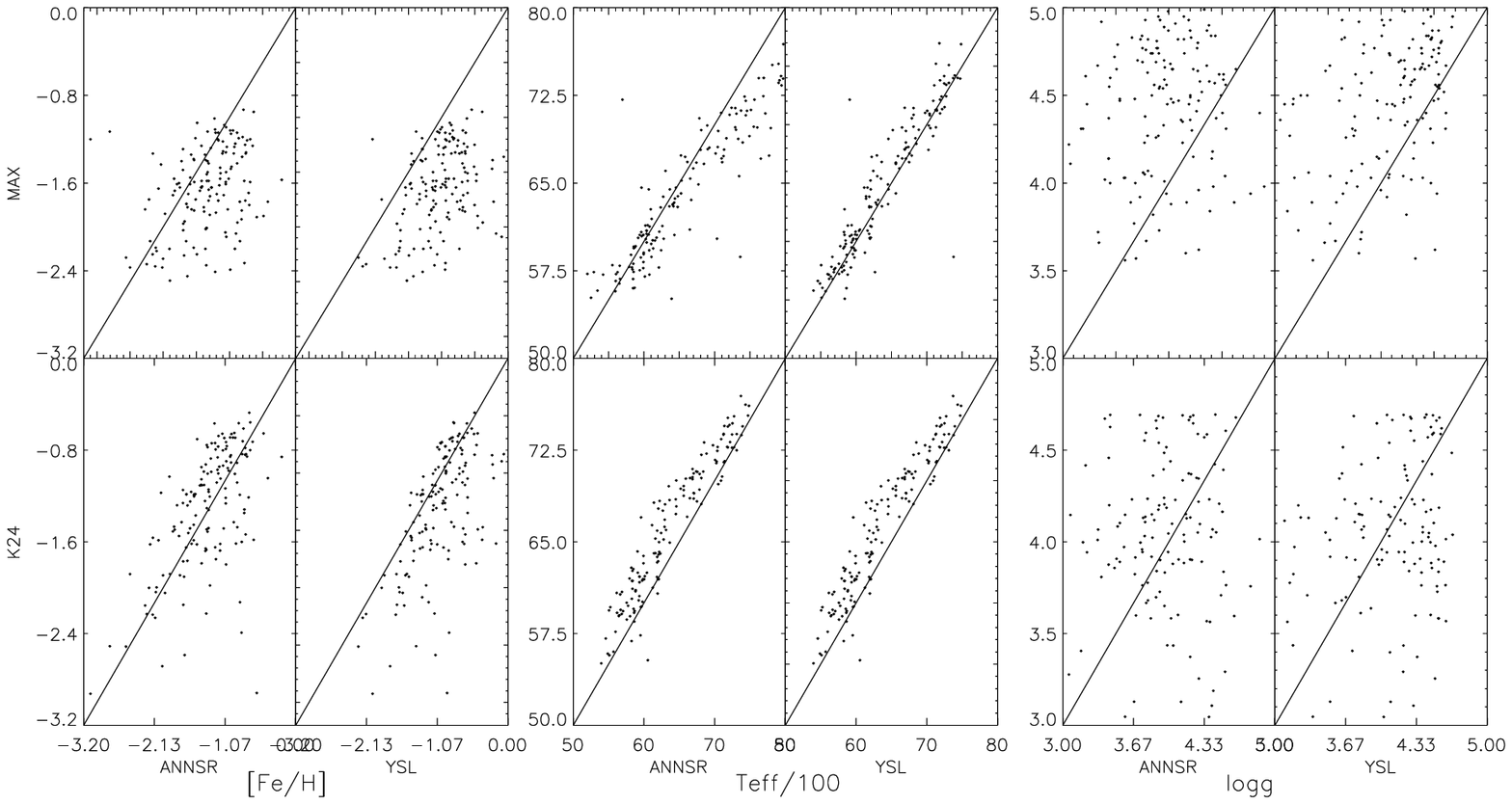}
\caption{Upper plot: Comparison of the results from the SEGUE Stellar Parameter Pipeline (SSPP) with different methods: This work (MAX, black), \citep[NGS1, blue]{lee08_1}, \citep[ANNSR, red]{re_fiorentin07} and \citep[k24, green]{ap06}. Lower part:  Comparisons of individual methods for a randomly selected subsample of stars. Each plot has a line with slope of unity. The different offsets and scatters between the results are indicated in Table~\ref{tab_corr}. In each plot, a selection of 150 random stars has been used. }
\label{corr_sspp}
\end{figure*}

\subsection{Comparison with the SEGUE Stellar Parameter Pipeline}

The SEGUE Stellar Parameter Pipeline \citep[SSPP,][]{lee08_1, lee08_2, SSPPIII} is a combination of different techniques to estimate the stellar parameters of SEGUE. Some of them are to fit the data to a grid of synthetic spectra like the k24 \citep{ap06}, and the k13, NGS1 and NGS2 ones \citep{lee08_1}. Another method are the ANNSR and ANNRR of \citet{re_fiorentin07}, which are an artificial neural network (ANN) trained to use a grid of synthetic (S) or real (R) spectra to estimate the parameters of real spectra (R). Other options are line indices \citep{wilhelm99} and the \caii K line index  \citep{beers99}.\\
We compare our results with the final adopted SSPP value (ADOP), and those of the k24, NGS1 and ANNSR grids of synthetic spectra. Each of these results can be found in the SSPP tables of the SDSS database. The grids of models are created using Kurucz stellar atmosphere models, but k24 and NGS1 cover the wavelength range of [4400,5500] {\AA}. The ANNSR grid includes the \caii triplet close to 8600 {\AA}. 

Figure~\ref{corr_sspp} shows the correlations between the results of these methods and our own. As in Fig.~\ref{corr_intrinsic}, we plotted a random selection of 150 stars to better visualize the correlations, but values of offsets and scatters were made using the entire sample of 17,274 stars.  The left panels correspond to the metallicity, the middle ones to temperature and the right ones to the surface gravity. The top row of plots shows  a general comparison of all the different methods,  plotted with different colors. The $y-$axis indicates the ADOP results and the $x-$axis the methods \Max~ (black), NGS1 (blue), ANNSR (red) and k24 (green). The one-to-one line is also plotted in the figures.  For each distribution a histogram of ($\mathrm{method_i - method_j}$) with $i \neq j$ was fitted with a Gaussian to obtain estimations for offset (mean $\Delta$) and scatter (standard deviation $\sigma$). The results are summarized in Table~\ref{tab_corr}. In the lower panel of Fig.~\ref{corr_sspp} we plotted the comparisons of individual methods for 150 randomly selected stars. These panels help to  better visualize how the single methods of the pipeline and our own correlate with each other. Figure~\ref{corr_sspp} and  Table~\ref{tab_corr} show that our results reasonably agree with the pipeline, and that the scatter is of the same order on magnitude as that between the individual SSPP methods. 

Considering individual parameters, our metallicities have a general tendency to be $\sim 0.3$~dex lower than adopted metallicity of SSPP (ADOP).  A similarly large offset exists between ANNSR and NGS1, with $-0.22$~dex. This could be due to the consideration of the \caii lines in the fit, which are not included in the NGS1 and k24 ones. ANNSR also includes \caii lines, the offset of -0.17 dex to our results is smaller than 1 $\sigma$ error. If the same lines are used in the fitting, no offset should be seen, as will be shown in Sect.~ \ref{hires}. 

The temperature, on the other hand, shows a small offset of -61 K with respect to the ADOP value of the pipeline. It is encouraging to see that we obtain the best agreement, except for ANNSR, which has no offset.  A large mean difference is found between k24 and NGS1, where the offset is 244~K. The scatter in the various comparisons varies from 100 K (NGS1 v/s ADOP) to  200K (k24 v/s ANNSR); our scatter of 112 K with respect to ADOP is one of the lowest values.  

Finally, $\log g$ shows the largest offset. We derived gravity values 0.51 dex higher than  the ADOP ones. The worst case is the comparison between our method and ANNSR with 0.63 dex.  Between the methods of the pipeline, the largest difference (0.27 dex) is found between ANNSR and NGS1. The best agreement we found for gravities are with NGS1, with an offset of 0.38 dex. The scatter for gravities varies from 0.23~dex (NGS1 v/s ADOP) to 0.48~dex  (ANNSR v/s k24 and \Max v/s k24).   A possible approach to correct our gravities would be to shift the zero point by $-0.38$~dex to agree with those of NGS1. We prefer to accept that gravity is our least constrained parameter, given the lack of sensitive features except for the wings of the \mgi b lines, and the noise in the spectra does not restrict $\log g$ effectively. In order to deal with this problem, k24  and NGS1 smooth the spectra to  half of the resolution, gaining signal-to-noise by this procedure. We remark that the k24 grid includes g-r colors, also to constrain the temperature. This automatically leads to different gravity values. A more extensive discussion of this aspect will be given in Sect.~\ref{addition}. The right panels of Fig.~\ref{corr_sspp} shows that all $\log g$ estimates are rather uncorrelated with each other with respect to the other ones. \\

\begin{table*}
\centering
\caption{Offsets ($\Delta$) and scatter ($\sigma$) of the differences (raw - column) between the methods: MA$\chi$ (this work). ADOP \citep{lee08_1}, NGS1 \citep{lee08_1}, ANNSR \citep{re_fiorentin07} and k24 \citep{ap06} for the entire sample of selected SEGUE stars.}
\label{tab_corr}
\begin{tabular}{c| cccc}
\hline \hline
  & ADOP &NGS1 & ANNSR & k24\\
   & $\Delta  / \sigma$ & $\Delta / \sigma$ & $\Delta /  \sigma$ & $\Delta /  \sigma$\\

[Fe/H] & & & & \\
\hline
\Max &   -0.32 / 0.23 &  -0.41 / 0.29&   -0.17 / 0.33 & -0.34 / 0.26\\
k24 &  0.04 / 0.14& -0.07 / 0.25&  0.17 / 0.27&\\
ANNSR&  -0.13 / 0.22 & -0.22 / 0.23 &\\
NGS1 & 0.08 / 0.17& & &\\
\hline
\\
$\mathrm{T_{eff}}$ & & & & \\
\hline
\Max &  -61 / 112 & 80 / 132 &  -76 / 170 & -159 / 160\\
k24 &  98 / 101 & 244 / 166 & 88 / 195& \\
ANNSR & 1 / 114  & 147 / 173  & &\\
NGS1 &  -142 / 93 & & & \\
\hline
\\
$\log g$ & & & \\
\hline
\Max &0.51 / 0.39 & 0.38 / 0.43 &0.63 / 0.47 &  0.41 / 0.48\\
k24 & 0.08 / 0.35 & -0.02 / 0.42 & 0.26 / 0.48 & \\
ANNSR &  -0.14 / 0.36 &  -0.27 / 0.45 & & \\
NGS1 & 0.09 / 0.23 & &\\
\end{tabular}
\end{table*}

\subsection{Systematic errors: Choice of fiducial model}

The $\mathbf{b}-$vectors are basically derivatives with respect to the parameters.With the assumption that the responses of the model to these derivatives are continuous, the choice of the fiducial model (FM) used to build the $\mathbf{b}-$vectors is free. This assumption, however, is only correct for certain regions of the parameter space.
For example, Balmer lines are good temperature indicators for stars no hotter than 8000K \citep{gray_book}, and for cold stars spectral lines are affected by molecular lines. Limiting our candidate stars to the range of 5000K to 8000K we know that we have a clean spectra with Balmer lines as effective temperature indicators. This allows us to choose a FM with any temperature in that  range to construct our weighting vector $\mathbf{b}_{\mathrm{Teff}}$, which will represent well the dependence of the model spectrum on temperature. Similar assumptions are made for the other two model parameters.

\begin{figure}
   \centering
   \includegraphics[scale=0.45,angle=90]{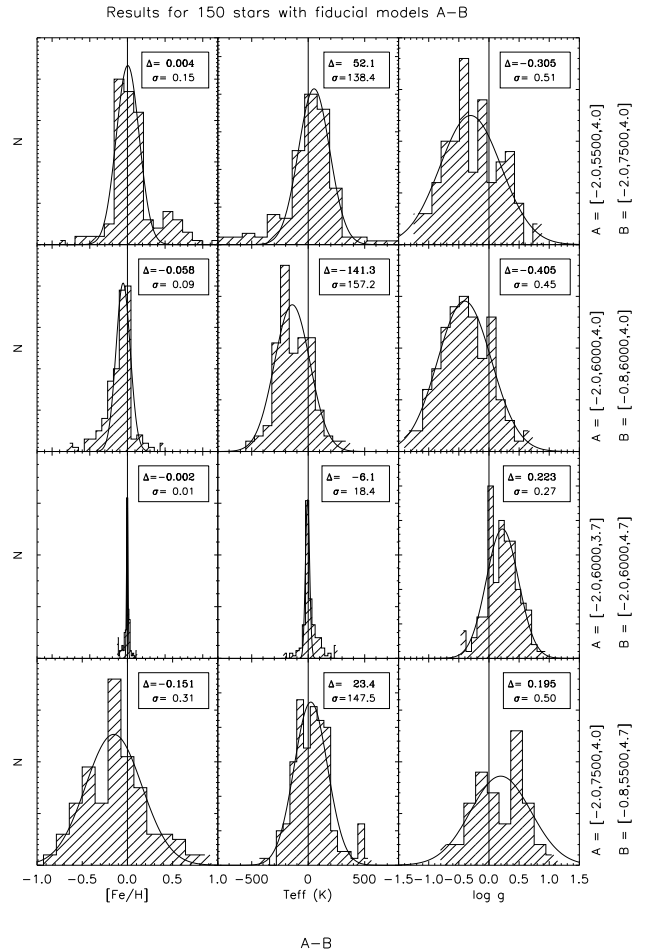}
   \caption{Histograms of the differences of the results obtained using 8 different fiducial models for the compression. Left: metallicity, middle: temperature and right: gravity. The parameters of the fiducial model are indicated in the right side of each plot, labeled as [[Fe/H], $\mathrm{T_{eff}}$, $\log g$]. Each histogram has a Gaussian fit, where the peak ($\Delta$) and the standard deviation ($\sigma$) are indicated in the  top legend.}
   \label{b_dependency}
   \end{figure}

In order to test this assumption we studied the effect of using different fiducial models in the final results of the parameter estimation. In Fig.~\ref{b_dependency} we plot the results of metallicity (left), temperature (middle) and gravity (right) for 150 randomly selected SEGUE stars. There are four sets of different fiducial models [A,B] used for the \textbf{b} vector calculation. The parameters of the FM are indicated at the right side in each set. From top to bottom the first one compares the  results of a cold ($\mathrm{T_{eff}} = 5500$ K) and a hot ($\mathrm{T_{eff}} = 7500$ K) fiducial  model of our grid of synthetic spectra, the second one  a metal-poor ($\mathrm{[Fe/H]} = -2.0$) and a metal-rich ($\mathrm{[Fe/H]} = -0.8$) one. The third one considers two different gravities ($\log g = 3.7$ and $\log g = 4.7$). For these three sets, all other parameters are identically calculated. Finally, the last one uses models which vary all three parameters. The histograms of each plot correspond to the difference of the results obtained with the fiducial model with a set of the parameters A and B, respectively. The Gaussian fit of the histogram is overplotted and its mean ($\Delta$) and standard deviation ($\sigma$) are indicated in the legend at the top. \\

Metallicity shows a  well defined behavior under different \textbf{b}-vectors. There are negligible offsets when varying only one parameter in the FM, except for the case where the FM is totally different. But even in that case the offset of -0.15 dex  is less than the 1 $\sigma$ errors of 0.25 dex obtained in the metallicity estimation. We conclude that this parameter does not show real systematic offsets due to the choice of fiducial model.  The scatter obtained in the metallicity is also less than the 1 $\sigma$ errors of the results. Only in the last case (bottom left) it becomes comparable with the errors. Because gravity is a poorly constrained parameter (see discussions above) and the $y-$numbers are uncorrelated, the effect when using different gravities in the FM should be completely negligible in the result of the other parameters. This can be seen in the third panel from top to bottom of Fig.~\ref{b_dependency}, where metallicity and temperature show extremely small offsets and scatters.

The derivation of temperature using different fiducial models shows a similar behavior. The discrepancies between results when using different FM are on the order of 50 K, except when the metallicity is varied. The mean difference of the final temperature estimates when using a metal-poor fiducial model and a metal-rich one is of 141 K, which is smaller than some differences seen in Table~\ref{tab_corr} between the methods of the SEGUE Stellar Parameter Pipeline.  This probably happens because the spectrum of a metal-rich star presents more lines than a metal-poor one, which is translated into the \textbf{b}-vector as temperature-dependent regions that do not exist. In the case of SEGUE data, many of the weak lines seen in a metal-rich synthetic spectrum and(or) the \textbf{b}-vector of a metal-rich fiducial model are hidden by the noise in the observed spectra and the assumption of a temperature dependence due to these weak lines may not be correct.  For the analysis of low signal-to-noise spectra, it is preferable to choose a FM with a rather low metallicity. But even when using a high metallicity for the FM, the offset is comparable with the 1 $\sigma$ error of 150 K for the temperature estimations. We conclude that temperature does not show a significant systematic offset due to choice of fiducial model either.

Finally, gravity shows larger discrepancies in the final results when varying the FM. The differences can be as large as 0.4 dex in the worst case when using different metallicities in the FM. This could also be because a metal-rich FM contains more lines and therefore greater sensitivities to gravity, which in our low signal-to-noise spectra is not the case. These sensitivities are confused with the noise in the observed spectrum. Scatters of 0.5 dex are in general on the order of the $1 \sigma$ errors.

\begin{figure*}
\centering
\includegraphics[width=20cm,angle=90]{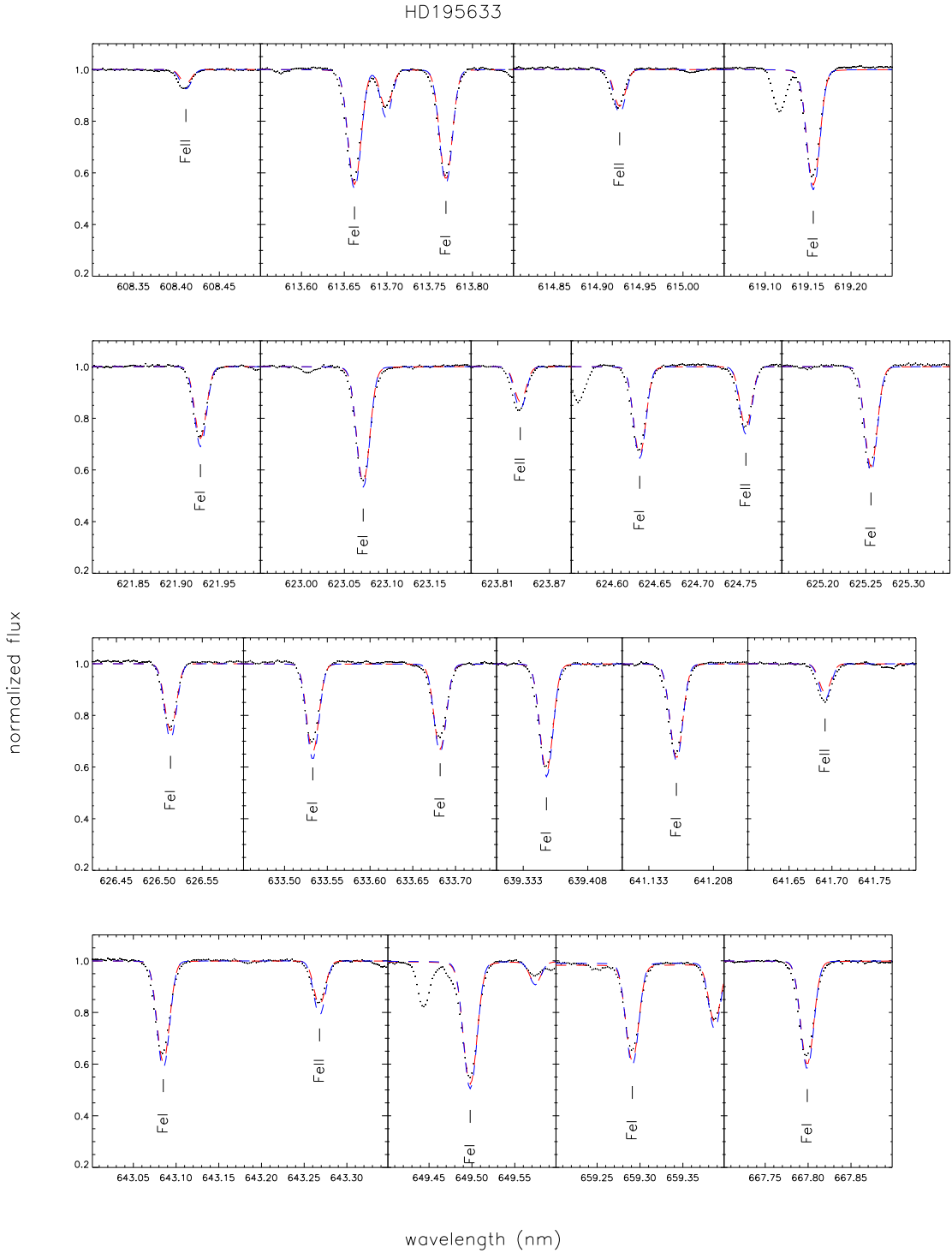}
\caption{Fit of the spectrum of the star HD195633 (points) with the best model (red dashed line). The parameters of the synthetic spectrum are [Fe/H] = -0.5, T$_{\mathrm{eff}} = 6318$ K and $\log g$ = 4.61, with $\chi^2 = 6.31$. The blue line corresponds to the model with  [Fe/H] = -0.6, T$_{\mathrm{eff}} = 6005$ K and $\log g$ = 3.86 ($\chi^2 = 7.31)$, obtained with the ``classical'' approach.  Each panel represents a spectral window where the lines from Table \ref{Fe_lines} are labeled at the bottom.}
\label{fit_uves}
\end{figure*}

%%%%%%%%%%%%%%%%%%%%%%%%%%%%%%%%%%%%%%%%%%%%%%%%%

\section{Application to high-resolution spectra}\label{hires}

As a check, MA$\chi$ also was tested with a smaller sample of 28 high-resolution spectra. For consistency with the low resolution
implementation, again  metal-poor dwarf stars were selected.

\subsection{Data}

The spectra were obtained with UVES, the Ultraviolet Echelle
Spectrograph \citep{dekker} at the ESO VLT $8\,m$ Kueyen telescope
in Chile. The resolving power is R $\sim$ 43,000 and the  signal-to-noise
is typically above 300 in our spectra. Because \Max~ is an automatic fitting tool
for synthetic spectra, we first had to be sure that the observed
spectra could be fitted by our grid of synthetic spectra. This means that we
had to have easily distinguishable unblended spectral lines and a
clear continuum. For these reasons we selected a part of the red
setting of the UVES spectra, which covers the wavelength range 580 nm
- 680 nm, where there are many unblended lines.

\subsection{Grid of high-resolution models}
When modeling spectra in high resolution, there are some differences
to the low-resolution case and the following considerations have to be
made:

\begin{itemize}
\item  A value for microturbulence must be set, because the shape of strong lines
 sensitively  depends on the value chosen for $v_{t}$. This is not the
  case in low resolution, so
  that there is no need to fine-tune this parameter.
  In high resolution the set of basic stellar
  atmosphere parameters is
  $([\mathrm{Fe/H}],\mathrm{T_{eff}},\log g, v_{t})$.
  To do a proper parameter estimation we
  would need to create a four-dimensional grid of synthetic spectra, varying all
  the parameters indicated above. This goes further than the purposes
  of this paper, where we aim to check the applicability of our
  method. Therefore we fixed the microturbulence parameter to a typical
  value, guided by the results of a standard ``classical'' spectral analysis (see
  below).
\item To compute synthetic spectra, a list of lines with their wavelength and atomic data must be
  included. The atomic data consist of oscillator strength $\log
  gf$ and excitation potentials, which are an important source of
  differences in the final shape of a given line. For the comparison
  consistent line lists are needed.
\item In order to make a proper fit to a broad wavelength range, the
  models should be perfect. Our models were computed under the
  simplifying assumption of LTE and with plane-parallel atmosphere
  layers; the atomic data have errors, too. There are too many features to fit, therefore a global fit
  becomes difficult, almost impossible. For this
  reason we selected windows of a limited spectral range and fitted the
  observed spectra only in these.
\end{itemize}

We determined the three parameters temperature, metallicity and gravity using
neutral and ionized iron lines. Neutral lines are sensitive to
temperature and single ionized ones to gravity \citep{fuhrmann98}. The metallicity was
effectively obtained from both Fe\small{I} \normalsize and \small{II}\normalsize, for a given temperature and
gravity. In
the wavelength range of 580nm - 680nm, we have six Fe\small{II} \normalsize lines and 20 Fe\small{I} \normalsize
ones that are unblended and strong enough to be present in most
metal-poor stars of our sample. The lines are indicated in
Table~\ref{Fe_lines}, where wavelength and $\log gf$ values are
listed. The $\log gf$ values and excitation potentials are taken from
\citet{nissen02} and the VALD database \citep{vald}.

Our grid of high resolution models covers a range in metallicity of $-2.5
\leq \mathrm{[Fe/H]} \leq -0.5$, in effective temperature of $5500
\leq \mathrm{T_{eff}} \leq 6600$, in surface gravity of $3.5 \leq
\log g \leq 5$ and in scaling factor for normalization from 0.85 to 1.15. The grid steps are the same
as in Sect.~\ref{grid_sdss}. The models have  $[\alpha/\mathrm{Fe}] = +0.4$ and $v_t = 1.2$ km/s, according to the values obtained for our stars with the ``classical'' method (see below).

\begin{table}[htdp]
\caption{Fe\small{I} \normalsize and \small II \normalsize lines used for the high-resolution fits. The
  first column indicates
  the wavelength; the $\log gf$ value is taken from the VALD database
  and is given in the second column.}
\begin{center}
\begin{tabular}{c c }
\hline
wavelength ($\AA$) & $\log gf$\\
\hline
Fe\small{I} \normalsize &  \\
\hline
6136.62&-1.410\\
6137.69&-1.403\\
6191.56&-1.416\\
6219.28&-2.448\\
6230.72&-1.281\\
6246.32&-0.877\\
6252.56&-1.767\\
6265.13&-2.540\\
6335.33&-2.177\\
6336.82&-0.856\\
6393.60&-1.576\\
6411.65&-0.717\\
6430.85&-1.946\\
6494.98&-1.239\\
6592.91&-1.473\\
6677.99&-1.418\\

\hline
Fe\small{II} \normalsize & \\
\hline

6084.11&-3.881\\
6149.26&-2.724\\
6238.39&-2.754\\
6247.56&-2.329\\
6416.92&-2.877\\
6432.68&-3.501\\
\hline
\end{tabular}
\end{center}
\label{Fe_lines}
\end{table}%

\subsection{Preparing the observed data for synthetic spectral
  fitting}

The reduction of the data was carried out in the same way as in
Sect.~\ref{preparation}, with the following difference in the normalization
procedure. At this resolution, neither the polynomial-fitting approach by
\citet{ap06}  nor IRAF\footnote{IRAF is distributed by the National Optical Observatory, which is operated by the Association of Universities of Research in Astronomy, Inc., under contract with the National Science Foundation.} were able to  automatically find a good
pseudo-continuum. Hence, the observed spectra were
normalized interactively using Midas \citep{midas}, and the synthetic spectra were computed
with normalized flux. The compression procedure was the same as
in Sect.~\ref{compression}. Compressing the grid of high-resolution
spectra takes more time than for the low-resolution ones, but this has to be
carried out only once to obtain a new grid of
$y-$numbers.

\subsection{Results of the high-resolution analysis using MA$\chi$}\label{uves}

Figure~\ref{fit_uves} shows an example fit of HD195633
(points) with the best MA$\chi$ model (red dashed line) and the model with
parameters found with the ``classical'' method (blue dashed line; see
below). Each panel
represents a spectral window with the lines of Table~\ref{Fe_lines}
used for the parameter estimation. All three parameters were
determined simultaneously and the best fit corresponds
to the model with parameters of [Fe/H] = -0.5, T$_{\mathrm{eff}} =
6318$ K and $\log g$ = 4.61. The final parameters of all stars in
this sample are given in Table~\ref{params}. The first column of the table
indicates the name of the star and the six next ones are the
parameters found with two different  MA$\chi$ analyses. The first set
(``free'') corresponds to the standard approach of determining all
parameters simultaneously from the spectrum only.
In the second variant (``restricted'') the gravity is determined
independently using Eq.~(14), which will be
introduced below. The last four columns are the
parameters obtained from the ``classical'' approach (below).

\small
\begin{table*}[htdp]
\caption{Parameters of the stars (Col~1) obtained with MA$\chi$ for
  both types of analysis discussed in the text. In the first
  analysis all three parameters are determined with only the
  spectrum, while in the second one $\log g$ was determined
  independently by using parallaxes according to Eq.~(14), as for the classical method (EW; last four columns)}
\begin{center}
\begin{tabular}{c| c c c| c c c| c c c c}
\hline \\

& \multicolumn{3}{c|}{MA$\chi$, free} & \multicolumn{3}{c|}{MA$\chi$,
  restr.} & \multicolumn{4}{c}{classical, EW} \\

star	&$T_\mathrm{eff}$ (K) & $\log g$ & [Fe/H]&$T_\mathrm{eff}$ (K)
& $\log g$ & [Fe/H]& $T_\mathrm{eff}$ (K) & $\log g$& [Fe/H]&$v_t$ (km/s)\\ \\

\hline \\
HD3567 & 6103.5& 4.79&-1.200 &6176.0 &4.20 &-1.223 & 6035.0 &  4.08& -1.200& 1.50\\
HD19445 & -- &--&--&5746.5 &4.10 &-2.293 & 5982.0  & 4.38 &-2.075 &1.40\\
HD22879 &5770.6&4.31&-0.877& 5778.1& 4.31& -0.880 & 5792.0&   4.29& -0.885 &1.20\\
HD25704& 5638.9 &3.61&-1.094 & 5864.4 &4.19 &-0.912  &5700.0   &4.18 &-1.010 &1.00\\
HD63077&5795.5&4.21&-0.796 &5787.7& 4.21& -0.793&  5629.0 &  4.15 &-0.935 &0.90\\
HD63598&6078.6&4.79&-0.681&5804.1 &4.23 &-0.901 & 5680.0   &4.16 &-0.885 &0.90\\
HD97320 &5601.1&4.65&-1.411&6025.1& 4.24 &-1.183 & 6105.0&   4.18 &-1.175 &1.20\\
HD103723 &5999.2&4.29&-0.896&6004.4 &4.29 &-0.896 & 6128.0  & 4.28 &-0.805 &1.50\\
HD105004 &5802.5&4.33&-0.900&5802.5& 4.33& -0.900 & 5900.0&   4.30& -0.800 &1.10\\
HD106038&5994.6&4.48&-1.392& 5995.5 &4.48 &-1.392  &5950.0   &4.33 &-1.390 &1.10\\
HD113679 &6037.3&4.73&-0.500&5757.5 &4.35& -0.708  &5759.0  & 4.04 &-0.590 &0.90\\
HD116064&5757.7&4.58&-2.41 &5757.7 &4.58 &-2.401  &5999.0   &4.33 &-2.000 &1.50\\
HD121004&5509.2&4.05& -0.670&5657.8 &4.40 &-0.650  &5711.0   &4.46 &-0.725 &0.65\\
HD122196 &5811.8&3.50&-1.954&5810.2 &3.56 &-1.954  &6048.0   &3.89 &-1.760 &1.20\\
HD126681&5556.9&4.44&-1.055 &5521.6 &4.68 &-1.075  &5532.0   &4.58 &-1.170 &0.60\\
HD160617 &5804.1&3.50&-1.961&5985.2 &3.75 &-1.794  &6028.0   &3.79 &-1.785 &1.30\\
HD175179 &6080.5&5.00&-0.500&5779.5 &4.47 &-0.691  &5758.0   &4.16 &-0.690 &0.90\\
HD188510 &5568.8&4.35&-1.341&5500.0 &4.65 &-1.231  &5536.0   &4.63 &-1.600 &1.00\\
HD189558 &5610.8&3.58&-1.191&5500.0 &3.75 &-1.181  &5712.0   &3.79 &-1.145 &1.20\\
HD195633 &6318.2&4.61&-0.500&6037.7 &3.99 &-0.687  &6005.0   &3.86 &-0.625 &1.40\\
HD205650 &5693.7&4.93&-1.195&5754.0 &4.47 &-1.208  &5842.0   &4.49 &-1.170 &0.90\\
HD298986 &5891.5&3.66&-1.592&6297.9 &4.22 &-1.304  &6144.0   &4.18 &-1.410 &1.40\\
CD-333337 &5808.2&3.66&-1.598&5946.9 &3.85 &-1.490 &5952.0 &3.95 &-1.435 &1.40\\
CD-453283 &5500.0&4.26&-0.978&5885.5 &5.00 &-0.685 &5657.0 &4.97 &-0.835 &0.80\\
CD-571633 &5939.9&4.37&-0.893&5939.9 &4.37 &-0.893 &5907.0 &4.26 &-0.930 &1.10\\
CD-3018140& 6143.1&3.50&-2.095&6246.5& 3.50& -1.980 &6340.0 &4.13 &-1.870 &1.00\\
G005-040 & 6228.2&5.00&-0.584&5930.4  &4.52 &-0.794 & 5766.0   &4.23 &-0.840 &0.80\\
G013-009 &6529.5&4.58&-2.29& 6556.0&  4.08 &-2.299 & 6416.0  & 3.95& -2.225 &1.40\\
\hline
\end{tabular}
\end{center}
\label{params}
\end{table*}%

\normalsize

\subsection{Results of a ``classical'' analysis}

In order to compare our MA$\chi$ results, we  analyzed the spectra
with a classical procedure, determining the stellar parameters through
an iterative process. The method is basically the same as in, e.g.,
\citet{nissen02}:

For the effective temperature it relies on the infrared
flux method, which provides the coefficients to convert colors to
effective temperatures. We used the (V-K) color and the calibration of
\citet{alonso96}, after converting the K 2MASS filter \citep{2mass}
to the Johnson filter \citep{bessel05} and de-reddening the
color. The extinction was taken from Schlegel's dust maps
\citep{schlegel} in the few cases where we could not find the values
in \citet{nissen02,nissen04}.

Surface  gravities were determined using the basic parallax relation
\begin{eqnarray}\label{logg}
\log{g} & = & \log{\frac{M}{M_{\odot}}} + \log{\frac{T}{T_{\odot}}} + 0.4~(Vc + BC)   \nonumber \\ 
&   &+ 2  \log{\pi} + 0.12 + \log {g_{\odot}},
%\end{center}
\end{eqnarray}

\noindent where $Vc$ is the $V$ magnitude corrected for interstellar
absorption, $BC$ the bolometric correction and $\pi$
the parallax in arcsec. We adopted a different mass value for each star based on those of  \citet{nissen02}, which are between 0.7 and 1.1 M$_{\odot}$. The bolometric correction $BC$ was calculated with the solar calibration of $BC_\odot$ = -0.12 as in \citet{Nissenlogg}.

The equivalent widths of neutral and ionized iron lines were used to
determine the metallicity [Fe/H]. It was
obtained using \tt{Fitline} \rm \citep{fitline}, which fits Gaussians to the line profiles. The
equivalent widths computed from these Gaussians fits were converted to
abundances by running MOOG
\citep[][Sobeck priv. comm.  2008]{moog} with the plane-parallel LTE
MARCS model atmospheres of \citep{marcs3}, which differ from those
used for MA$\chi$.  The line list contains the lines of
Table~\ref{Fe_lines}  and more taken from bluer wavelengths in the range
300 - 580 nm (Hansen, priv. comm. 2009).

Finally, the value for the microturbulence was found by
requiring that all the equivalent widths of neutral lines should give
the same Fe abundance as the ionized lines.

In order to obtain the four
parameters in this way, we started with an initial
guess for each of the interdependent parameters. After determining
them with the above steps all values will change due
to the interdependence, hence we had to iteratively determine the parameters until their values showed a negligible change.

We are aware that our metallicity based on Fe\small I \normalsize lines is ignoring any non-LTE
effects \citep{asplund05,collet05}. But because neither the
classically nor the MA$\chi$
analysis are taking this into account, the results
obtained by both methods can well be compared.

The upper panels of Fig.~\ref{corr_uves} show the results of metallicity (left),
temperature (middle) and gravity (right) for the entire sample of
stars. The $x-$axis shows the results of the automatic fitting
MA$\chi$ and the $y-$axis the ``classical'' parameter estimation (EW).
The one-to-one line is overplotted in each figure and the legend
indicates the mean of the differences (MA$\chi$ - EW) and its standard
deviation, denoted as offset and scatter, respectively. Each point
corresponds to a star of Table~\ref{params} with parameters obtained by
the ``free'' method.

Gravity shows a scatter of $\sigma = 0.46$
dex and a negligible offset. The determination of the gravity from
FeII lines has always been a problematic task: \citet{fuhrmann98} in an
extensive study of parameters of nearby stars showed that surface
gravities of F-type stars located at the turn-off point can easily
differ by up to $\sim$ 0.4~dex if derived from either LTE iron
ionization equilibrium or parallaxes.   The amount  of ionized lines (gravity dependent)  present in the
spectra is usually smaller than that of neutral ones (temperature dependent). By performing an
automatic fitting of weighted spectra that contain only six Fe\small{II} \normalsize lines
compared to 20 Fe\small I \normalsize lines, a scatter of 0.46 dex is reasonable.

The result for the effective temperature shows a very small offset of
only $-10$~K, but a quite
large scatter of $\sigma = 227$ K. Given that two
different methods were employed to determine it, this  still appears to
be acceptable.  The scatter may also be
affected by our fixed value for the microturbulence, which was taken
from the average of the values found with the EW approach (see
Table~\ref{params}), but which in individual cases may differ severely.

One can decrease the scatter in the temperature difference by removing
the three most discrepant objects from the sample. The \textbf{star} HD63598 (triangle) has an
offset of 398~K. The Schlegel maps give a de-reddening for this star that is unrealistically large, so instead we set it to zero. This led to a temperature that was too low. \citet{hd63598} have found a temperature of 5845~K for this star, reducing the difference to 233~K with respect to our result.
The star G005-040
(asterisk) is the second case, where the offset is 462~K. For this
star, the continuum subtraction was not perfect in every spectral
window, resulting in a fit where the parameters were quite
unreliable. Finally, the weak lines of the observed spectrum of
HD19445
pushed MA$\chi$ to the border of the grid. The parameters in this case
were undetermined.
By removing these three stars from our sample, the
scatter for temperature is reduced to 197~K.

Metallicity, on the
other hand, shows a very good agreement, with a negligible offset of
0.02~dex and a scatter of 0.16~dex (upper left panel of Fig.~\ref{corr_uves}). The
lines used for the automatic fitting (MA$\chi$) and for the classical
analysis (EW) are in most cases identical, except for the lowest metallicities,
where some of them are hardly visible. In these cases the classical method
 also resorted to other lines outside the MA$\chi$ wavelength range.
The atomic data are identical, but the value for microturbulence are not, as
mentioned before. In view of all this, the very good correlation of metallicity
is encouraging.

\bf
\subsection{``Restricted'' parameter recovery}\label{addition}
\begin{figure*}
\centering
\includegraphics[width=7cm,angle=90]{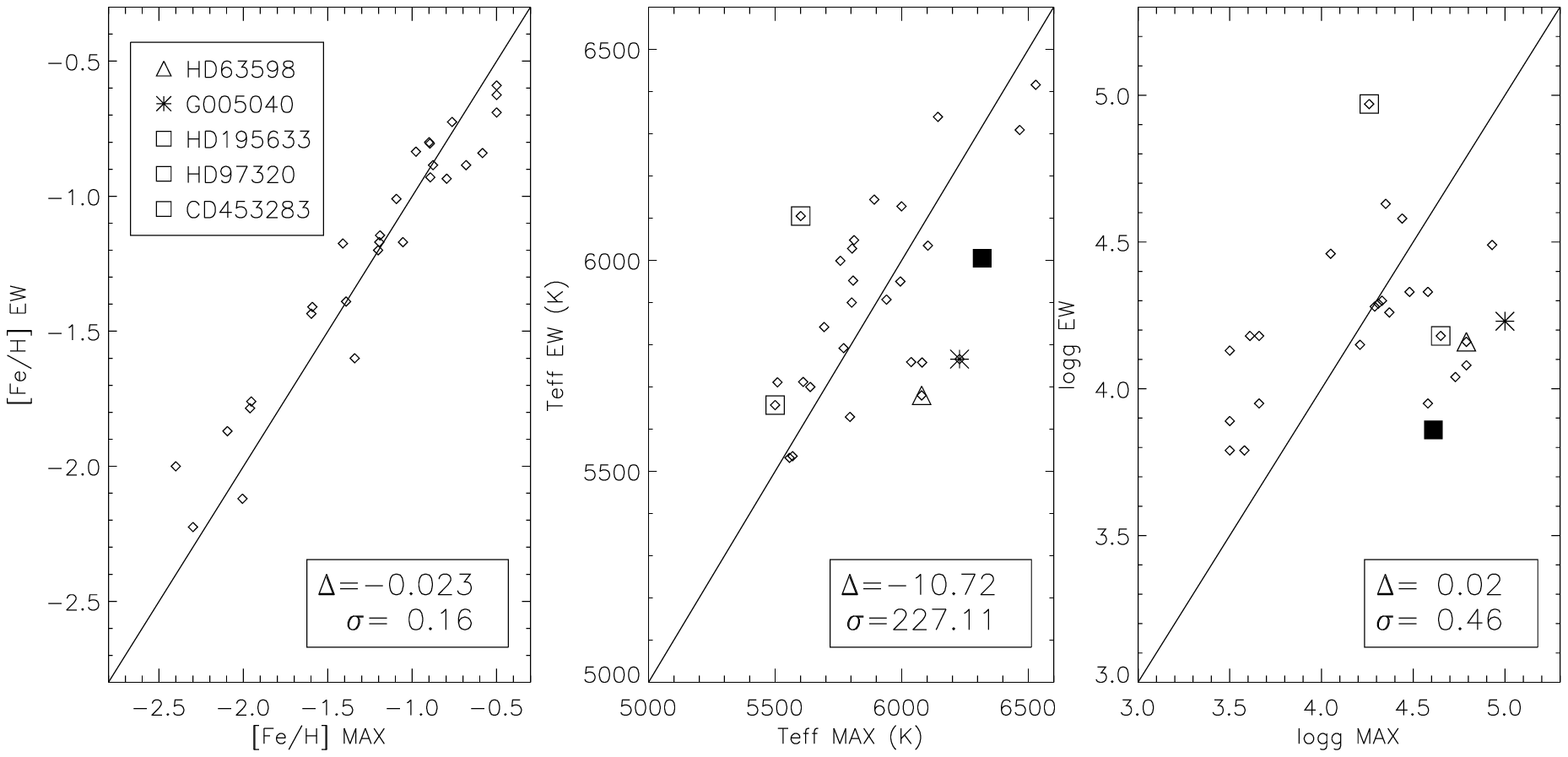}
\includegraphics[width=7cm,angle=90]{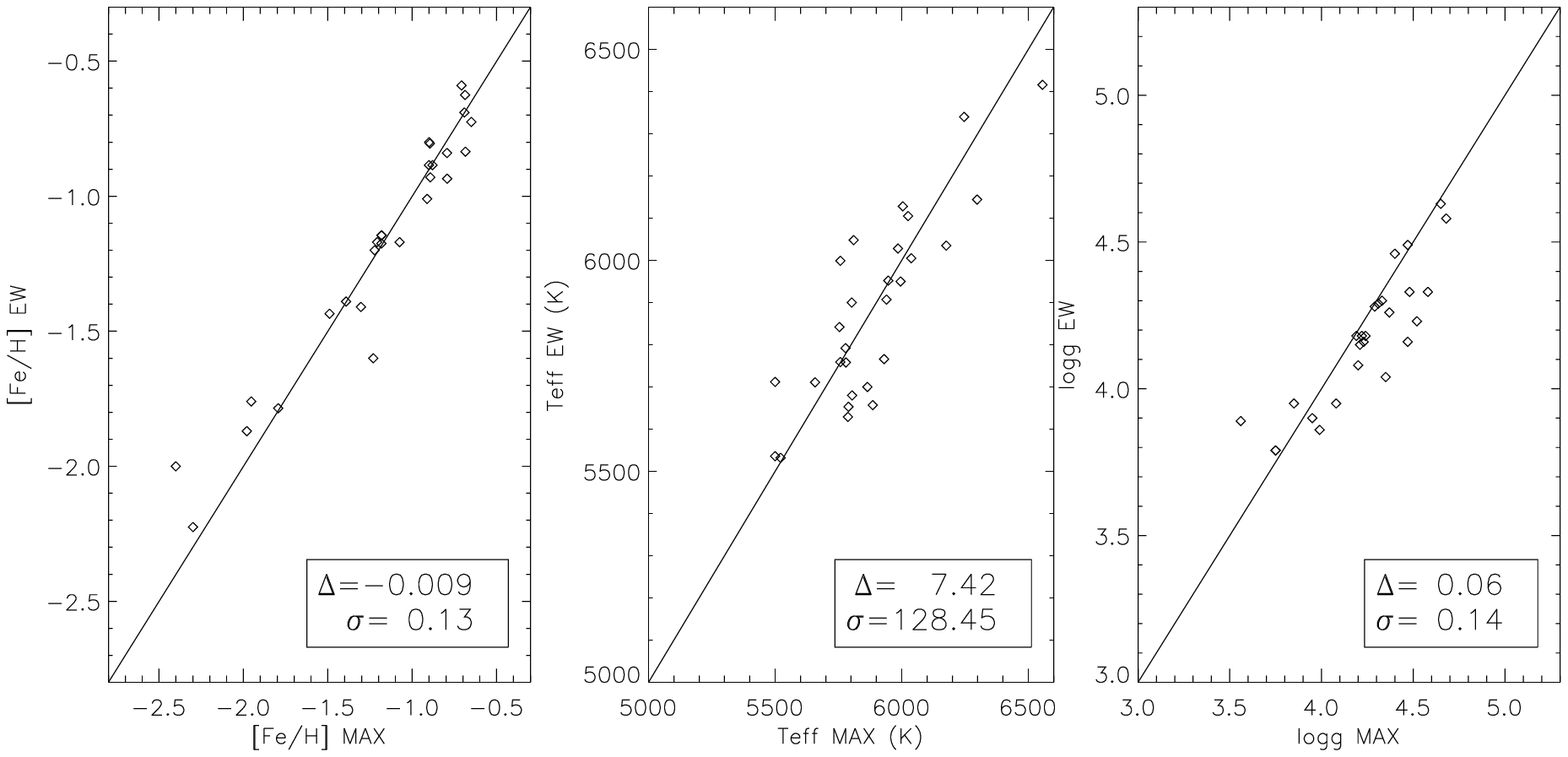}
\caption{Upper panels:  Correlations between the results obtained with the automatic
  fitting (MAX) and the classical method (EW) for metallicity (left),
  temperature (middle) and gravity (right). In each panel, the
  one-to-one line is overplotted and a legend containing the mean difference (in the sense
  ``EW-MA$\chi$'') and its standard deviation.  Here the three parameters were determined simultaneously
  and directly from the spectra by
  our  method. The stars with special symbols are indicated in the legend and correspond to special cases where the offset between the two methods is particularly large (discussed in  text). 
 Lower panels:  As upper panels, but MA$\chi$ results were obtained
  forcing the method to find the best-fitting model close to the
  gravities computed from the parallax formula, Eq.~(14). Individual
  data can be found in Table~\ref{params} (columns ``MA$\chi$, free'' for the upper case and ``MA$\chi$, restr.'' for the lower one). }
  \label{corr_uves}
\end{figure*}

\rm
An example for the individual fit of the best MA$\chi$ model to the
observed spectra was shown in Fig.~\ref{fit_uves} for HD195633,
which is shown as a filled symbol in Fig.~\ref{corr_uves} and is one
of the objects with the most pronounced 
discrepancies (see also Table~\ref{params}). Nevertheless, the two
synthetic spectra appear to fit equally well the
observations ($\chi^2 = 6.31$ for the MA$\chi$ model and $\chi^2 = 7.31$ for that
obtained from the classical analysis).  We could not decide from the $\chi^2$ values which
method leads to a more accurate parameter determination for this
star. One would need independent information of
the spectrum about this star to reach conclusions about its
parameters.

It is also interesting to notice from the upper panels of Fig.~\ref{corr_uves} that stars
with large discrepancies in gravity
(for example those with special symbols) also give large discrepancies in temperature, as expected. They show
a quite unsatisfying fit for the Fe\small{II} \normalsize lines. Motivated by this, we
did an additional test by restricting MA$\chi$ to the determination
of the parameters with input values for
$\log g$ obtained from the EW method, i.e. we found a local maximum point of the likelihood in a restricted area. 
To do this we chose the three closest gravity values from our grid of synthetic spectra to the classical one found with Eq.~(\ref{logg}) -- see Col~9 of Table~\ref{params} -- and we searched for the maximum point within this range. May be that for $\log g$ there is no local maximum in this range, and the final estimation will go to the border of the grid, as the case of CD-3018140, where $\log g = 3.5$. The general tendency is anyway a local maximum close to the input EW-value.

Now the agreement
with the classical method for the effective temperatures became excellent, with a negligible offset of only
7.42 K and a scatter of 128 K, as seen in the lower panels of Fig.~\ref{corr_uves}.
Gravity was also better constrained, with a small scatter of 0.14 dex
and a negligible offset. Note that we did not necessarily obtain
identical values, because the MA$\chi$ gravity is obtained by using the value from Eq.~(\ref{logg}) only as input, and we were looking for a final solution close to this value. The behavior of the metallicity
does not change with respect to the ``free'' case, demonstrating the
robustness of the determination of this parameter.  The results
obtained when using parallaxes for the initial guess for  $\log g$ are
summarized in Table~\ref{params} in the columns under the heading
``MA$\chi$, restr.''.

It is instructive to discuss the implication of this comparison.
For the ``classical'' EW method we determined the
parameters making use of the best information available and of the
freedom to adopt the method to each star individually. The
iterative process allowed us to decide where to stop the iteration, or,
if no satisfying convergence could be reached, to draw on the options to
move  to another spectral window, to use other lines, or disregard
problematic lines. Moreover, the continuum could be
separately subtracted for each line, thus creating locally
perfect normalized flux levels. For the effective temperature, the
more reliable infrared flux method employing photometric data could be used
instead of only relying on spectra. The advantage of determining $\log g$ independently was
already demonstrated.

On the other hand we were
attempting an estimate of the parameters only from the spectral
information without fine-tuning the models or fit procedure for each
star for our automatic MA$\chi$ method,. Given this, the comparison with the full interactive method is
surprisingly good. We demonstrated that by using additional information
for one parameter (here $\log g$) it becomes even better for the
remaining two quantities.  Table~\ref{params} shows our final parameter
estimate for our stars, when we make use of the parallaxes as additional
information.

The result of this test is that we demonstrated our ability to
estimate the basic stellar atmosphere parameters quickly and accurately
enough for a substantial sample of stars. While individually severe outliers
may occur,  the method is accurate overall for a whole population of stars. For this purpose the EW method would be much too
slow and tedious.
Our method may also serve as a quick and rough estimate for a more
detailed follow-up analysis.

%%%%%%%%%%%%%%%%%%%%%%%%%%%%%%%%%%%%%%%%%%%%%%%%

\section{Summary and conclusions}\label{fin}

We  described \Max, a new derivative of \moped for the estimation of parameters from stellar spectra.   In this case the parameters were the metallicity, effective temperature and surface gravity.  The method reduces the data to a compressed data set of three numbers, one per parameter. Assuming that the noise is independent of the parameters,  the compressed data contain as much information about the parameters of the spectrum as the entire data itself. As a result, the likelihood surface around the peak is locally identical for the entire and compressed data sets, and the compression is 'lossless'. This massive compression, with the degree of compression given by the ratio of the size of the data to the number of parameters, allows the cost function calculation for a parameter set to be sped up by the same factor. For SDSS data, the spectra are of the order of 1000 datapoints, each with a corresponding error, which must be compared to a similarly sized model. The speed up factor in this case is then at least $1000/3$, $\sim333$x.

This extremely fast multiple parameter estimate make the \Max~ method a powerful tool for the analysis of large samples of stellar spectra. We have applied it to a sample of 17,274 metal-poor dwarf stars with low-resolution spectra from SEGUE, using  a grid of synthetic spectra  with the parameter range of $[-2.5, -0.5]$ dex in metallicity, [5000,8000] K in effective temperature and [3.5, 5] dex in surface gravity, covering a wavelength range of [3850, 5200] \AA.

From the \caii, Balmer and \mgi b lines, which are the strongest absorption features identified in SDSS spectra, we estimated the metallicity with averaged accuracies of 0.24 dex, the temperature with 130 K and $\log g$ with 0.5 dex, corresponding to the 1$\sigma$ errors. Surface gravity is a poorly constrained parameter using these data, mainly due to the lack of sensitive features of this parameter (apart from some degree of sensitivity of the wings of the \mgi b triplet) and the considerable noise in the spectra when compared to high-resolution spectra. Additional information to the spectra, such as photometry, would help to constrain the gravity parameter more. 

MA$\chi$ has the option to simultaneously fit different spectral windows. We have compared estimates of the parameters using the whole spectrum and only those data ranges where known lines exist. Both analyses take approximately the same time and agree excellent in recovered parameters. This suggests that for these low-resolution spectra there is no need to carefully and laboriously select the spectral windows to be analyzed with MA$\chi$, the method calculates this automatically as part of its weighting procedure. 

With the assumption that the spectra behave similarly under changes of the parameters, the choice of the fiducial model for the compression is free. We have created eight compressed grids using different fiducial models for the $\mathbf{b}-$vector calculation and obtain agreement between the recovered parameters. Given the low signal-to-noise of our data, it is better to use a more metal-poor fiducial model, as the dependence on the parameters will be focussed on the strong lines. 

We have comprehensively investigated the correlations of our results with those obtained for the SEGUE Stellar Parameter Pipeline \citep{lee08_1}, which reports results from a number of different methods. The results from MA$\chi$ agree well with those of the various pipelines, and any differences are consistent with those between the various accepted approaches. We are aware of the  MATISSE \citep{matisse} method of parameter estimate, which uses a different combination of weighting data but is closer to the MA$\chi$ approach than the standard pipeline methods. We look forward to comparing  our results with those of MATISSE when they become available.

More specifically, temperature agree excellently with the averaged temperature of SSPP, with a negligible offset of -61 K and low scatter of 112 K. Our metallicities show a tendency to be -0.32 dex lower than SSPP averages. The small scatter of $0.23$ dex suggests that the different spectral features used in the analysis (mainly Ca\small{II} \normalsize lines) could shift the zero point of the metallicity. The most pronounced discrepancy is found in surface gravity, where Ma$\chi$ reports values 0.51 dex ($\sigma$ = 0.39) higher than the averaged value of the pipeline. We saw that this is consistent with the discrepancies seen between other methods.

We have also tested MA$\chi$ on a sample of 28 high resolution spectra from VLT-UVES, where no offset in metallicity is seen. In this case we have carefully chosen the models and spectral range for comparing our parameter estimation against a ``classical'' approach:   temperature from photometry, surface gravity from parallaxes and metallicity from equivalent widths of neutral and ionized iron lines. We have calculated the parameters  ourselves to avoid additional systematic offsets that various different methods would introduce in stellar parameter scales. These results were compared with the automatic fitting of 20 Fe\small{I} \normalsize and 6 Fe\small{II} \normalsize lines (that coincide in most cases with the equivalent widths calculations) made with MA$\chi$. We obtained large scatter in gravity and temperature (0.44 dex and 220 K, respectively), but no systematic offset. The normalization of the continuum in some cases was not perfectly done, making it difficult to fit every line properly: especially for the Fe\small{II} \normalsize lines. They produce a scatter in gravity, which drives a scatter in temperature.\\
We have seen in our fits that our best model does not differ very much from the model with parameters found by the ``classical'' approach. Motivated by this  we decreased the scatter in gravity by using additional information to the spectrum, i.e. using the gravities determined from parallaxes as input value. This forced our method to find a local maximum point of the likelihood close to this input gravity value. Fixing the gravity gave an improved agreement in temperature, now with a scatter of 128 K.  Metallicity does not change when forcing gravities, illustrating the robustness of determination for this parameter. 

MA$\chi$ is  an extremely rapid fitting technique and as such is independent of model and data used. It will work for any star for which an appropriate grid of synthetic spectra can be calculated. Although the grid calculation is time consuming, it only needs to be performed once allowing  an extremely rapid processing of individual stars.

MA$\chi$ is one of the fastest approaches to parameter determination, and its accuracy is comparable with other methods. To develop MA$\chi$ further in preparation for the next generation of surveys (eg. Gaia, APOGEE or LAMOST), we plan to expand our range of model grid and integrate photometric data to improve the determination of the parameters, especially surface gravity.  We must be aware that in a grid of synthetic spectra with a broader parameter range a set of  \textbf{b}-vectors from one single fiducial model will not represent well  the dependence on the parameters in the entire sample. To cover the entire sample with well represented dependences, we plan to compute different compressed grids, each of them with \textbf{b}-vectors calculated from different fiducial models. The parameter estimate in this case will be made with the different compressed grids using a final convergence test. After convergence, the final parameters should be estimated from the compressed grid with the fiducial model with parameters close to the final result.

\begin{acknowledgements}
This work is part of the PhD thesis of Paula Jofr\'e and is funded by an IMPRS fellowship. We  thank Richard Gray, Martin Asplund, Francesca Primas and Jennifer Sobeck for their helpful comments in interpreting our results, Alan Heavens for algorithmic advice and especially Carlos Allende Prieto, for all the great advice and for sharing with the authors the continuum subtraction routine. P. Jofr\'e is thankful to Timo Anguita and Manuel Aravena for programing tips and Thomas M\"adler for his careful reading of the manuscript and the support to take this paper out. Finally, the authors gratefully acknowledge the many and detailed positive contributions made by the referee. 
Funding for the SDSS and SDSS-II has been provided by the Alfred P. Sloan Foundation, the Participating Institutions, the National Science Foundation, the U.S. Department of Energy, the National Aeronautics and Space Administration, the Japanese Monbukagakusho, the Max Planck Society, and the Higher Education Funding Council for England. The SDSS Web Site is http://www.sdss.org/.\\
The SDSS is managed by the Astrophysical Research Consortium for the Participating Institutions. The Participating Institutions are the American Museum of Natural History, Astrophysical Institute Potsdam, University of Basel, University of Cambridge, Case Western Reserve University, University of Chicago, Drexel University, Fermilab, the Institute for Advanced Study, the Japan Participation Group, Johns Hopkins University, the Joint Institute for Nuclear Astrophysics, the Kavli Institute for Particle Astrophysics and Cosmology, the Korean Scientist Group, the Chinese Academy of Sciences (LAMOST), Los Alamos National Laboratory, the Max-Planck-Institute for Astronomy (MPIA), the Max-Planck-Institute for Astrophysics (MPA), New Mexico State University, Ohio State University, University of Pittsburgh, University of Portsmouth, Princeton University, the United States Naval Observatory, and the University of Washington.

\end{acknowledgements}
\bibliographystyle{aa} %%%%%%%%%%%%%%%%%%%%%%%%%%%%%%%%%%%%%%%%%%%%%%%%%%
\bibliography{max_paper}
 % p_galaxies_bib.bib
\clearpage
\end{document}